\documentclass[sigconf]{acmart}
\usepackage{fancyhdr}
\usepackage{amsmath,amssymb,amsfonts}
\usepackage{algorithmic}
\usepackage{graphicx}
\usepackage{textcomp}
\usepackage{amsthm}
\usepackage{xcolor}
\usepackage[ruled,linesnumbered,boxed]{algorithm2e}
\usepackage{bm}
\usepackage{bbm}
\usepackage{dsfont}
\usepackage{array}
\usepackage{subfigure}
\usepackage{enumitem}
\usepackage{multirow}
\usepackage{multicol}
\usepackage{tabularx}
\newcommand{\tabincell}[2]{\begin{tabular}{@{}#1@{}}#2\end{tabular}}

\AtBeginDocument{%
  \providecommand\BibTeX{{%
    \normalfont B\kern-0.5em{\scshape i\kern-0.25em b}\kern-0.8em\TeX}}}
    
\renewcommand\footnotetextcopyrightpermission[1]{}

\makeatletter

\newcommand{\Rmnum}[1]{\expandafter\@slowromancap\romannumeral #1@}
\makeatother

\setcopyright{acmcopyright}
\copyrightyear{2018}
\acmYear{2018}
\acmDOI{10.1145/1122445.1122456}

\acmConference[Woodstock '18]{Woodstock '18: ACM Symposium on Neural
  Gaze Detection}{June 03--05, 2018}{Woodstock, NY}
\acmBooktitle{Woodstock '18: ACM Symposium on Neural Gaze Detection,
  June 03--05, 2018, Woodstock, NY}
\acmPrice{15.00}
\acmISBN{978-1-4503-XXXX-X/18/06}

\begin{document}

\title{Disentangled Graph Contrastive Learning for Review-based Recommendation}

\author{Yuyang Ren}
 \affiliation{
   \institution{Shanghai Jiao Tong University}
 }
\email{renyuyang@sjtu.edu.cn}

\author{Haonan Zhang}
 \affiliation{
   \institution{Shanghai Jiao Tong University}
 }
\email{zhanghaonan@sjtu.edu.cn}

\author{Qi Li}
 \affiliation{
   \institution{Shanghai Jiao Tong University}
 }
\email{liqilcn@sjtu.edu.cn}

\author{Luoyi Fu}
 \affiliation{
   \institution{Shanghai Jiao Tong University}
 }
\email{yiluofu@sjtu.edu.cn}

\author{Jiaxin Ding}
 \affiliation{
   \institution{Shanghai Jiao Tong University}
 }
\email{jiaxinding@sjtu.edu.cn}

\author{Xinde Cao}
 \affiliation{
   \institution{Shanghai Jiao Tong University}
 }
\email{xdcaosjtu@outlook.com}

\author{Xinbing Wang}
 \affiliation{
   \institution{Shanghai Jiao Tong University}
 }
\email{xwang8@sjtu.edu.cn}

\author{Chenghu Zhou}
 \affiliation{
   \institution{Institute of Geographical Sciences and Natural Resources Research, Chinese Academy of Sciences}
 }
\email{zhouch@lreis.ac.cn}

\begin{abstract}
User review data is helpful in 
alleviating the data sparsity problem in many recommender systems. In review-based recommendation methods, review data is considered as auxiliary information that can improve the quality of learned user/item or interaction representations for the user rating prediction task. However, these methods usually model user-item interactions in a holistic manner and neglect the entanglement of the latent factors behind them, e.g., price, quality, or appearance, resulting in suboptimal representations and reducing interpretability. In this paper, we propose a Disentangled Graph Contrastive Learning framework for Review-based recommendation (DGCLR), to separately model the user-item interactions based on different latent factors through the textual review data. To this end, we first model the distributions of interactions over latent factors from both semantic information in review data and structural information in user-item graph data, forming several factor graphs. Then a factorized message passing mechanism is designed to learn disentangled user/item representations on the factor graphs, which enable us to further characterize the interactions and adaptively combine the predicted ratings from multiple factors via a devised attention mechanism. Finally, we set two factor-wise contrastive learning objectives to alleviate the sparsity issue and model the user/item and interaction features pertinent
to each factor more accurately. Empirical results over five
benchmark datasets validate the superiority of 
DGCLR over the state-of-the-art methods. Further analysis is offered to interpret the learned intent factors and rating prediction in DGCLR.

\end{abstract}

\vspace{-0.1cm}
\vspace{-0.1cm}

\vspace{-0.1cm}


\maketitle
\vspace{-0.2cm}
\section{Introduction}
Review-based recommendation aims to alleviate the data sparsity problem \cite{mao2016multirelational,shi2019deep} in collaborative filtering
(CF) methods \cite{koren2022advances} by using the reviews of users to items as auxiliary information. Textual reviews contain useful semantic information that can be associated with the basis of users for their ratings, thereby leading 
to several investigations \cite{zheng2017joint,chen2018neural,xi2021deep} aimed to evaluate these user reviews to improve user preference modeling and rating predictions.

Traditional review-based recommendation methods usually employ
topic models, e.g., Latent Dirichlet Allocation (LDA) \cite{blei2003latent} and word embedding model \cite{mikolov2013efficient} to learn latent feature distributions of users and items \cite{wang2011collaborative,mcauley2013hidden,pena2020combining}. 
As deep learning rapidly develops, deep neural networks such as 
Convolutional Neural Network (CNN) and Recurrent
Neural Network (RNN) are introduced to model the review data for 
recommendation \cite{zheng2017joint,wang2015collaborative,catherine2017transnets}. Moreover, motivated by the attention mechanism \cite{vaswani2017attention}, many attention-based methods \cite{chen2018neural,wu2019context,wu2019hierarchical} are  proposed to identify the different importance of components, such as sentences, reviews, and users/items for better recommendation. More recently, the success of graph neural works \cite{kipf2016semi,velivckovic2017graph} in modeling the graph data also inspires its application in review-based recommender systems \cite{wu2019reviews,shuai2022review,gao2020set}. Since 
the user-item interactions can be naturally represented as a graph, review signals are also incorporated into the graph learning process to learn user/item embeddings.

\begin{figure}\setlength{\abovecaptionskip}{-0.05cm}
\centering
\centering
\includegraphics[width=2.7in]{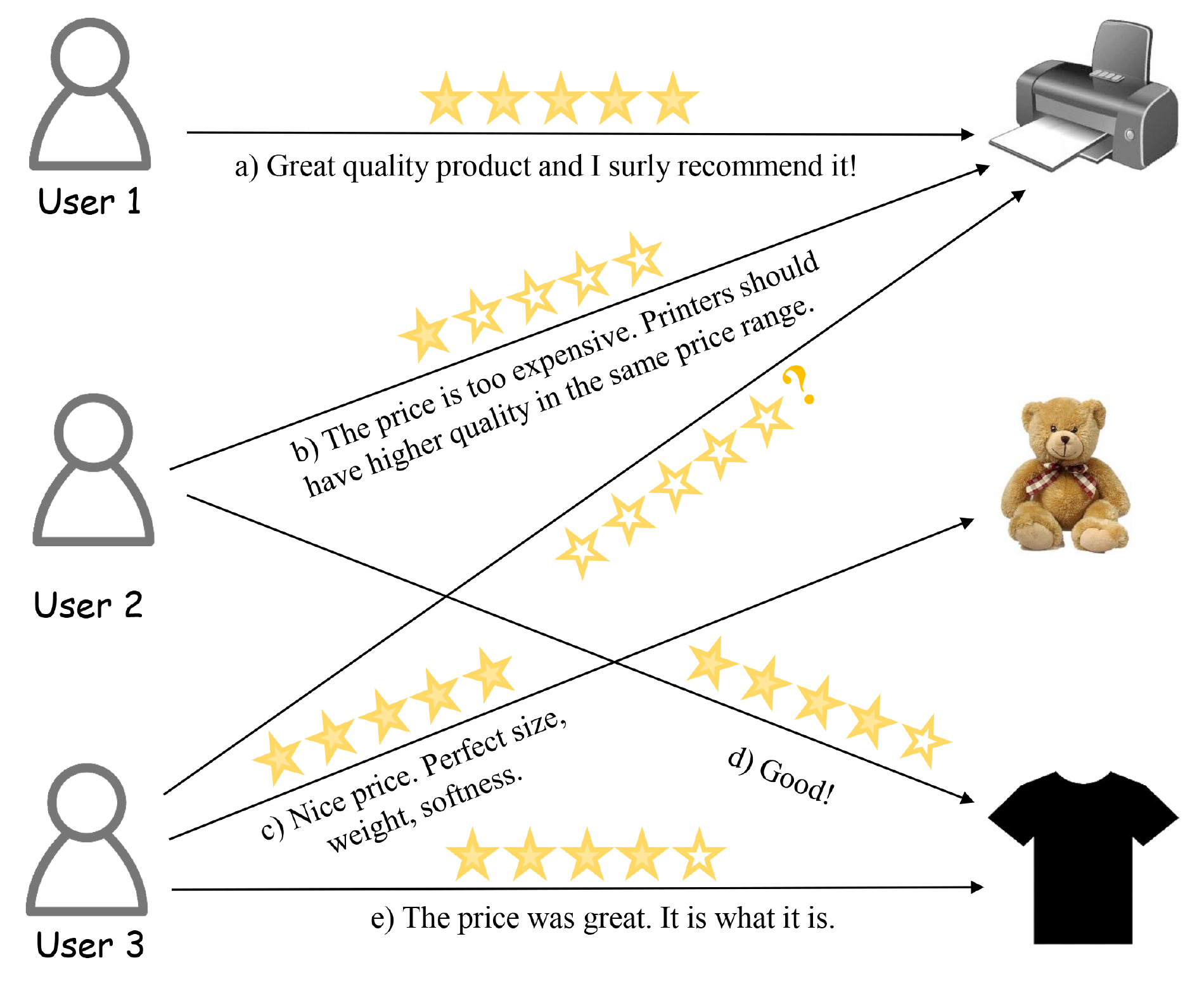}
\centering
\caption{An example of user-item rating graph.} 
\label{fig1}
\vspace{-0.7cm}
\end{figure}

Despite their success, existing methods typically employ a holistic approach to leverage review data for user preference or user-item interaction modeling, i.e., either aggregating user/item reviews for user/item embedding learning \cite{chen2018neural,zheng2017joint,wu2019reviews} or approximating the review of each user-item interaction based on learned user/item embeddings \cite{catherine2017transnets,sun2020dual,xi2021deep}. Recently proposed RGCL \cite{shuai2022review} moves them forward by modeling the reviews as edge features in the user-item graph and incorporating them into the message passing process. However, user ratings of items are typically influenced by various complex latent factors, such as price, quality, appearance, etc. As shown in Figure \ref{fig1}, User 1 likes the printer because of its excellent quality, whereas User 2 dislikes it for its cheap cost performance. When it comes to predicting User 3's rating to the printer, we notice that User 3 is price sensitive based on his/her interactions with other items, so we anticipate a low rating score. Therefore, the complex latent factors underlying user-item interactions highlight a desire to disentangle these factors in the review-based recommendation, which is still unexplored. As a result,
the user/item and interaction representations learned by 
existing methods contain a jumble of entangled factors, reducing interpretability and leading to suboptimal recommendation performance.

In this paper, we propose to learn disentangled user/item and interaction representations for better and more explainable review-based recommendation. To this end, we borrow the idea from disentangled representation learning (DRL) \cite{higgins2016beta}, which aims to learn factorized representations to characterize the latent factors hidden in the data. Although introduced for some other recommendation tasks \cite{wang2020disentangled,ma2020disentangled,chen2021curriculum}, DRL in review-based recommendation faces the following challenges. 
\setlist[itemize]{leftmargin=*}
\begin{itemize}

    \item How to accurately identify the distribution of latent factors in the user-item interactions based on review and graph information and model the user preference at a finer granularity?
    \item How to characterize the interactions from multiple factors and distinguish the decisive factor for rating prediction?
    \item How to design proper self-supervised tasks based on the factorized representations of users, items, and interactions to alleviate the sparsity issue and encourage disentanglement?

\end{itemize}

To tackle these challenges, we propose a novel Disentangled Graph Contrastive Learning framework for Review-based recommendation (DGCLR). In particular, we first design a disentangled graph learning (DGL) module equipped with graph disentangling and factorized message passing mechanisms. The former models the distribution of latent factors in each user-item interaction jointly from semantic information in the review and structure information in the user-item graph. The latter characterizes the user preferences from various aspects based on the generated factor graphs by accumulating factor-relevant information from neighborhoods. Then an attention-based interaction (AI) module is created to learn the factorized interaction representations  and combine the predicted ratings from various latent factors. Furthermore, we present two factor-wise contrastive learning (CL) tasks that generate self-supervised signals for model learning. Specifically, we design a factor-wise node discrimination (FND) task to enhance user/item embeddings and a factor-wise edge discrimination (FED) task to align factorized interaction features with review information from different aspects. This design encourages the learned representations to be disentangled and better model the user/item characteristics pertinent to each latent factor. In comparison to existing methods, DGCLR learns disentangled representations for users, items, and interactions,
allowing it to investigate the meaning of each latent factor, resulting in greater explainability for predicting user ratings.

Our contributions can be mainly summarized as:
\begin{itemize}
    \item We design a DGL module and an AI module  to learn disentangled representations for users/items and interactions at a finer granularity.
    \item We propose a novel DGCLR framework based on two factor-wise contrastive learning tasks that encourage disentanglement as well as alleviate the data sparsity issue.
    \item We conduct extensive experiments on five real-world datasets to validate the effectiveness and interpretability of DGCLR.
\vspace{-0.1cm}
\end{itemize}

\vspace{-0.2cm}
\section{Related Work}
\vspace{-0.1cm}
\subsection{Review-based Recommendation}
In the early stages of research, topic models are usually employed to obtain latent feature distributions of users and items from the review data. For example, the CTR model \cite{wang2011collaborative} represented items with the sum of
topic factors and free embeddings. The HFT model \cite{mcauley2013hidden} combined latent rating
dimensions with latent review topics learned by LDA \cite{blei2003latent}. In the TIM model \cite{pena2020combining}, topic factors were used to initialize both user and item embeddings.

As deep learning develops, many advanced text methods are used to extract semantic information from the review data. For instance, Wang et al. \cite{wang2015collaborative} proposed collaborative deep learning, which jointly employs Stacked Denoising AutoEncoders to learn representations for the content information and probabilistic matrix factorization to learn users’ rating behaviors. DeepCoNN \cite{zheng2017joint} employed two parallel
TextCNNs \cite{2014Convolutional} to extract semantic features from reviews. To tackle the
problem of unavailable target reviews at the inference stage, TransNet \cite{catherine2017transnets} extended DeepCoNN by introducing an additional layer to approximate the review of a target user-item pair. Following the approximation strategy, some works were also proposed, such as DualPC \cite{sun2020dual} and DRRNN \cite{xi2021deep}.

In addition, attention mechanism \cite{vaswani2017attention} is introduced due to its ability to identify the key elements.
NARRE \cite{chen2018neural} learned the
review representation through CNN, and scored each review
through the attention mechanism. CARL \cite{wu2019context} employed CNN with attention mechanism to highlight the relevant semantic information by jointly considering the reviews written by/for a user/item. DAML \cite{liu2019daml} employed dual attention mutual learning to model the importance of reviews and  integrated the rating features and review features into a neural network for rating prediction.

Graph neural networks (GNNs) \cite{kipf2016semi,velivckovic2017graph} extend deep learning techniques to process the graph data and have been also used for review-based recommendation. For example, RMG \cite{wu2019reviews} applied a three-level
attention network to learn representations of sentence, review, and user/item and a graph attention network to model
interactions. RGCL \cite{shuai2022review} incorporated review information as edge features into user/item embedding learning and designed two contrastive learning tasks as additional self-supervised signals.

We inherit the idea of modeling users, items, and reviews as an edge-feature enhanced graph and learn  disentangled user/item and interaction representations  with factor-wise self-supervised signals for better and more interpretable recommendation. 
 \begin{figure*}\setlength{\abovecaptionskip}{-0.05cm}
\centering
\centering
\includegraphics[width=\linewidth]{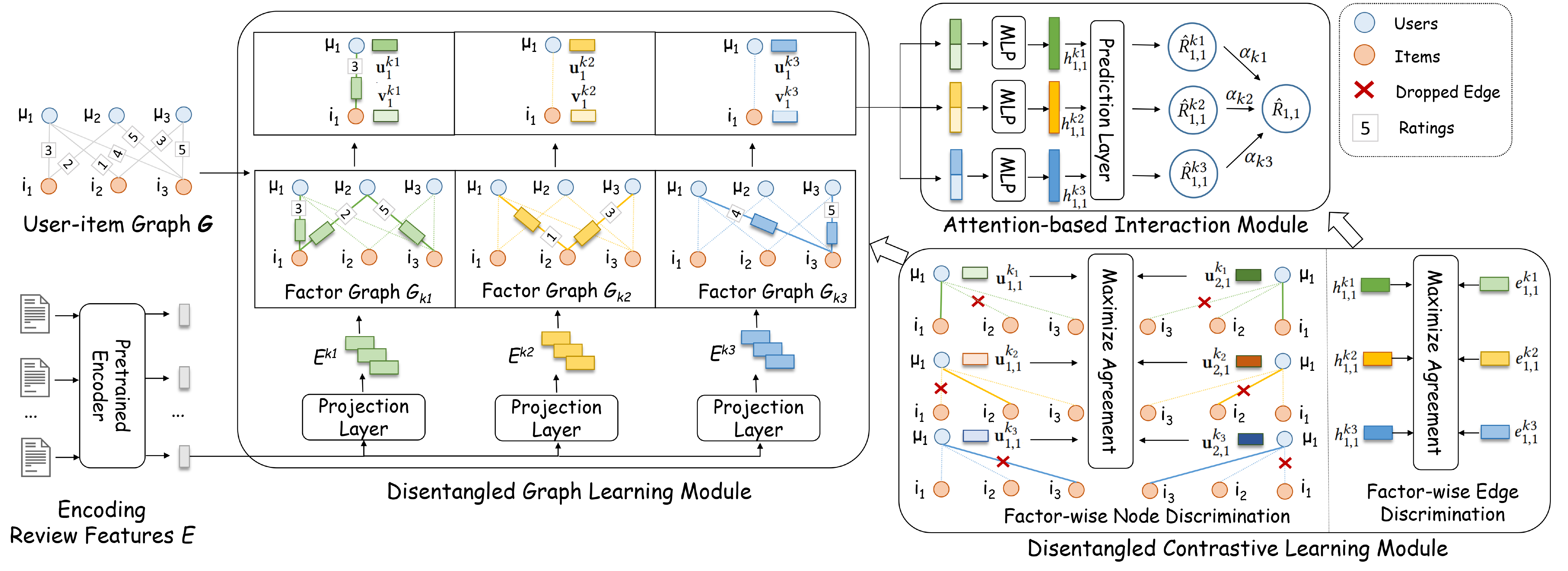}
\centering
\caption{Overview of DGCLR. For clarity, we show the pipeline
to generate the rating prediction for one user-item interaction.} 
\label{fig2}
\vspace{-0.4cm}
\end{figure*}
\vspace{-0.2cm}
\subsection{CL-based Recommendation}
In recent years, inspired by the success of contrastive learning (CL) in self-supervised representation learning \cite{oord2018representation,chen2020simple,you2020graph}, many CL paradigms are designed to alleviate the data sparsity issue \cite{yu2018adaptive} and boot the performance in recommender systems. For instance, SGL \cite{wu2021self} utilized an auxiliary GCL task to enhance user/item representation learning via self-discrimination. SEPT \cite{yu2021socially} mined multiple positive samples with semi-supervised learning on the perturbed graph for social-based recommendation. Zhang et al. \cite{zhang2021double} leveraged hypergraph to model user/item data, and proposed a double-scale node dropout strategy to generate self-supervised signals for
group recommendation. While these works focus
on designing CL tasks by regarding user/item representations as a whole, we design two factor-wise CL tasks to force the factorized user/item embeddings independently reveal user/item's properties.

\vspace{-0.2cm}
\subsection{Disentanglement-based Recommendation}
Disentangled representation learning aims to learn factorized
representations that reveal and disentangle the underlying latent factors hidden in the observed
data \cite{ma2019disentangled, yang2020factorizable}. When it comes to recommendation, MacridVAE \cite{ma2019learning} employed beta-VAE \cite{higgins2016beta}
on interaction data and achieved macro disentanglement
by inferring the high-level concepts associated with user behaviors. DGCF \cite{wang2020disentangled} factorized the user-item graph into several intent-aware
interaction graphs and iteratively update them based on user-item interaction. Ma et al. \cite{ma2020disentangled} proposed a sequence-to-sequence training strategy based on latent self-supervision and disentanglement for sequential recommendation. Despite the
promising performance, existing methods do not fit our task since they ignore
 the fruitful semantic information hidden in the review texts. Different from them, we learn the disentangled item/user and interaction representations from both semantic and structural aspects.
 \vspace{-0.1cm}
 \section{Problem Definition}
 In the task of review-based recommendation, we denote $\mathcal{U}$($|\mathcal{U}|=M$) as the user set and $\mathcal{I}$($|\mathcal{I}|=N$) as the item set. The rating record is formulated as a user-item rating matrix $R\in \mathbb{R}^{M\times N}$, where $R_{ij}$ denotes the rating score of user $i$ to item $j$ and $\mathcal{R}$ denotes the set of all the possible ratings in the dataset (e.g., $\mathcal{R}= \{1, 2, 3, 4, 5\}$  in Amazon).
 Meanwhile, the review texts are pre-processed to a fixed-length tensor $E \in \mathbb{R}^{M\times N\times d}$, where $\mathbf{e}_{i,j}$ denotes the feature of review text user $i$ comments on item $j$.
 Then the user-item interactions $\mathcal{E}$ can be represented by the combination of the rating matrix and the review tensor, i.e., $\mathcal{E}=(R, E)$. Finally, the review-based recommendation data can be formulated as a user-item bipartite graph $G=(\mathcal{U}\cup \mathcal{I}, \mathcal{E})$. The task is to predict the values of the full rating matrix $\hat{R}\in \mathbb{R}^{M\times N}$ based on the graph $G$.
\vspace{-0.1cm}
 \section{Proposed Model}
 This section gives a detailed introduction to our proposed  DGCLR. The overview of DGCLR is shown in Figure \ref{fig2}, which is composed of three parts: 1) Disentangled Graph Learning (DGL) Module: factorizing the input graph based on the user-item interactions and learning disentangled representations for users/items. 2) Attention-based Interaction (AI) Module: learning factorized interaction representations and predicting the rating matrix adaptively from multiple latent factors. 3) Disentangled Contrastive Learning (DCL) Module: introducing two auxiliary factor-wise CL tasks for the above two modules to alleviate the sparsity issue and encourage disentanglement. Our demonstration is unfolded as follows.
 \vspace{-0.2cm}
 \subsection{Disentangled Graph Learning Module}
 To model user/item's attributes pertinent to latent factors, we design a GNN model that learns disentangled representations for users/items. Each GNN layer consists of a graph disentangling mechanism which can accurately identify the latent factors in interactions to form multiple factor graphs, and a factorized message passing mechanism that performs multi-channel message passing on the factor graphs to aggregate factor-relevant features separately in each channel. Finally, multiple GNN layers are stacked to gather useful information from higher-order neighborhoods.
 
\vspace{-0.2cm}
 \subsubsection{Initialization}
 We follow \cite{shuai2022review,wu2019reviews} to
 parameterize user/item ID embeddings as free embedding
matrices $U\in \mathbb{R}^{M\times d}$ and $V\in \mathbb{R}^{N\times d}$. We further divide the ID embedding into $K$ chunks for separate user/item representation learning in each channel. Specifically, the ID embedding for user $i$ is represented as:
\begin{equation}
\vspace{-0.2cm}
    \mathbf{u}_{i}^{(0)} = (\mathbf{u}_{i}^{1, (0)},\mathbf{u}_{i}^{2, (0)},\dots,\mathbf{u}_{i}^{K, (0)})
\end{equation}
where $\mathbf{u}_{i}^{k,(0)} \in \mathbb{R}^{\frac{d}{K}}$ is user $i$'s chunked embedding of the $k$-th latent factor. Analogously, $\mathbf{v}_j^{(0)} = (\mathbf{v}_{j}^{1, (0)},\mathbf{v}_{j}^{2, (0)},\dots,\mathbf{v}_{j}^{K, (0)})$ is initialized as the ID embedding for item $j$.

 For review representations, we follow \cite{shuai2022review, hyun2018review} to encode user $i$'s review on item $j$ into the vector $\mathbf{e}_{i,j}$  with BERT-Whitening \cite{su2021whitening}, whose parameters are frozen during the model training for time and space efficiency considerations. Then we extract review's different features corresponding to the $K$ factors by projecting the review
 vector $\mathbf{e}_{i,j}$ into $K$ different subspaces:
 \begin{equation}
 \vspace{-0.1cm}
    \mathbf{e}_{i,j}^{k}=\sigma\left(\mathbf{W}_{k}^{\top} \mathbf{e}_{i,j}+\mathbf{b}_{k}\right)
\vspace{-0.1cm}
\end{equation}
where $W_k\in \mathbb{R}^{d\times \frac{d}{K}}$ and $\mathbf{b_k}\in \mathbb{R}^\frac{d}{K}$ are the parameters in the $k$-th channel, and $\sigma(\cdot)$ is a nonlinear activation function. 
We then assume that $\mathbf{e}_{i,j}^{k}$ approximately captures the aspect of review that is associated with the $k$-th factor, if $\mathbf{e}_{i,j}$ does contain relevant information about the related aspect.
\vspace{-0.2cm}
 \subsubsection{Graph Disentangling Layer}
After initialization, we then learn disentangled user/item representations via graph disentangling and factorized message passing.

\textbf{Graph Disentangling.} 
Considering the
fruitful semantic information, we propose to mine the distribution of latent factors from review texts. Specifically, given the review embedding $e^{k}_{i,j}$ of user $i$ to item $j$ in channel $k$, we present a prototype-based method to obtain the semantic score $\mathbf{se}_{i,j}^k$ that indicates how relevant is the review $(i,j)$ to factor $k$. We introduce $K$ latent factor prototypes ${\{\mathbf{c}_k\}}_{k=1}^K$ and the
score $\mathbf{se}_{i,j}^k$ is calculated as:
\begin{equation}
\vspace{-0.2cm}
    \mathbf{se}_{i,j}^{k}=\frac{\exp (\phi\left(\mathbf{e}_{i, j}^k, \mathbf{c}_{k}\right)/\tau)}{\sum_{k'=1}^{K} \exp (\phi\left(\mathbf{e}_{i, j}^{k'}, \mathbf{c}_{k'}\right)/\tau)}
\end{equation}
where $\phi(\cdot,\cdot)$ denotes the cosine similarity function and $\tau$ is the temperature hyperparameter.

Although the reviews can provide us some hints as to which factor user-item interactions fall into, there might be some missing information. Recalling the example in Figure \ref{fig1}, the review text (d) is general and cannot explicitly reflect the reason for user's rating. To tackle this problem, we propose to infer it from the neighborhood of user/item. Generally, if user $i$/item $j$ frequently interacts with its neighboring items/users based on factor $k$, we can draw the inference that user $i$ might also rating item $j$ based on factor $k$ with high probability. On the basis of this insight, we further introduce the  similarity  between user $i$ and item $j$ in terms of aspect $k$ to assist in judging the latent factors of interactions, which can be formulated as:
\begin{equation}
\vspace{-0.1cm}
    \mathbf{st}_{i,j}^{k, (l)}=\frac{\exp (\phi\left(\mathbf{u}_{i}^{k, (l-1)}, \mathbf{v}_{j}^{k, (l-1)}\right)/\tau)}{\sum_{k'=1}^{K} \exp (\phi\left(\mathbf{u}_{i}^{k', (l-1)}, \mathbf{v}_{j}^{k', (l-1)}\right)/\tau)}
\vspace{-0.1cm}
\end{equation}
where $\mathbf{st}_{i,j}^{k, (l)}$ denotes the structural score of user $i$ and item $j$ on the $k$-th factor at the $l$-th layer and $\mathbf{u}_{i}^{k, (l-1)}$/$\mathbf{v}_{j}^{k, (l-1)}$ denotes the learned embedding of user $i$/item $j$ at the $(l-1)$-th layer in the $k$-th channel. When $l=1$, $\mathbf{st}^{k, (l)}$ reflects the matching degree of user and item's own attributes on factor $k$; When $l>1$, $\mathbf{st}^{k, (l)}$ can integrate the information on factor $k$ from a larger receptive field due to the iterative accumulation of factor-relevant information from neighborhoods via factorized message passing.

Having modeled the distributions of latent factors from both semantic and structural perspectives, we then combine them into the final score $\mathbf{s}_{i,j}^{k, (l)}$  representing the coefficient of the edge between user $i$ and item $j$ in the $k$-th factor graph:
\begin{equation}
\vspace{-0.1cm}
    \mathbf{s}_{i,j}^{k, (l)}=\eta \mathbf{se}_{i,j}^{k}+ (1-\eta)\mathbf{st}_{i,j}^{k, (l)}
\vspace{-0.1cm}
\end{equation}
where $\eta\in[0,1]$ can be a hyperparameter or a learnable parameter. We empirically find that our model achieves a good performance in general when setting $\eta$ as $0.7$. This finding is consistent with our assumption that semantic information plays a dominant role in discriminating the factors in user-item interactions. As a result, we have derived a factor graph based on Equation (5) in each channel.

\textbf{Factorized Message Passing.} 
Given the learned factor graphs, we aim to leverage message passing to accumulate factor-relevant information for user/item representation learning. Specifically, we perform embedding propagation \cite{kipf2016semi} in each channel, such that the information of reviews and neighboring items/users, which are relevant to the factor,
are integrated into the learned user/item representations. 
Following \cite{berg2017graph,shuai2022review}, we treat rating score as edge type. Then for rating $r$, the factorized message passing from item $j$ to user $i$ in the $l$-th layer is formulated as:
\begin{equation}
\vspace{-0.1cm}
    \mathbf{x}_{r ; j \rightarrow i}^{k,(l)}=\frac{\mathbf{s}_{i,j}^{k,(l)} (\mathbf{e}_{i j}^{k}\cdot \mathbf{W}_{r}^{k,(l)}+ \mathbf{v}_{j}^{k,(l-1)})}
    {\sqrt{\left|\mathcal{D}_{j}^{k, (l)}\right|\left|\mathcal{D}_{i}^{k, (l)}\right|}},
\vspace{-0.1cm}
\end{equation}
where $W_{r}^{k,(l)}\in \mathbb{R}^{\frac{d}{K}\times \frac{d}{K}}$ is the parameter matrix to project the review embedding to the space of user/item embedding in the $k$-th channel. $\mathcal{D}_{i}^{k, (l)}=\sum_{p\in \mathcal{N}(i)}\mathbf{s}_{i,p}^{k,(l)}$ and $\mathcal{D}_{j}^{k, (l)}=\sum_{p\in \mathcal{N}(j)}\mathbf{s}_{p,j}^{k,(l)}$ denote the degrees of user $i$ and item $j$ in the $l$-th layer of channel $k$. Similarly, we can obtain the message passing from user $i$ to item $j$.

To intuitively figure out the essence of Equation (6), we hypothesize that factor $k$ represents \textit{price}. Then the interpretation is three-fold: 1) the coefficient $\mathbf{s}_{i,j}^{k,(l)}$ is capable of filtering out the noise information of reviews and items with which user $i$ do not interact due to price. 2) The review information $\mathbf{e}_{i,j}^{k}$ of user $i$ is collected to characterize his/her reviewing behaviors based on price. 3) The neighboring item feature $\mathbf{v}_{j}^{k,(l-1)}$ of user $i$ is accumulated to depict his/her price-sensitive preference on items.

After message passing in each channel, we then employ an aggregation operation similar to GC-MC \cite{berg2017graph} to aggregate all the factor-relevant messages, which is formulated as:
\begin{equation*}
    \mathbf{u}_{i}^{k,(l)}=\mathbf{W}^{(l)} \sum_{r \in \mathcal{R}} \sum_{p \in \mathcal{N}_{i, r}} \mathbf{x}_{r ; p \rightarrow i}^{k, (l)}, \quad \mathbf{v}_{j}^{k, (l)}=\mathbf{W}^{(l)} \sum_{r \in \mathcal{R}} \sum_{p \in \mathcal{N}_{j, r}} \mathbf{x}_{r ; p \rightarrow j}^{k, (l)}
\vspace{-0.1cm}
\end{equation*}
where $\mathbf{W}^{(l)}\in \mathbb{R}^{\frac{d}{K}\times \frac{d}{K}}$ is the parameter matrix and $\mathcal{N}_{i, r}$ is 
the set of items that user $i$ rates with rating $r$.
\vspace{-0.2cm}
\subsubsection{Layer Combination}
As mentioned before, our model benefits from the rich semantics of higher-order relationships by disentangling the user-item graph into the factor graphs. For example, the second-order connectivity $u_{1}^k \rightarrow i_{2}^k \rightarrow u_{3}^k$ indicates the intent similarity between $u_{1}$ and $u_{3}$ when rating $i_{2}$ based on the $k$-th factor.
To capture the useful information from higher-order neighbors, we further stack $L$ graph disentangling layers to form the final representations for users/items in each channel:
\begin{equation}
\vspace{-0.1cm}
    \mathbf{u}_{i}^k=\frac{1}{L}\sum_{l=1}^{L} \mathbf{u}_{i}^{k, (l)} ; \quad \mathbf{v}_{j}^k=\frac{1}{L}\sum_{l=1}^{L} \mathbf{v}_{j}^{k, (l)}
 \vspace{-0.1cm}
\end{equation}

\subsection{Attention-based Interaction Module}
As analyzed before, different users may rating different items based on different factors. In order to model the interactions between users and items from each latent factor, we concatenate the user and item embeddings in each channel and use a Multi-Layer Perceptron (MLP) to obtain the factorized interaction feature $h_{i,j}^k$:
\begin{equation}
\vspace{-0.1cm}
    \mathbf{h}_{i j}^k=\operatorname{MLP}\left(\left[\mathbf{u}_{i}^k, \mathbf{v}_{j}^k\right]\right),
\vspace{-0.1cm}
\end{equation}
where $\mathbf{h}_{i,j}^k \in \mathbb{R}^\frac{d}{K}$ denotes the learned interaction feature of factor $k$. The rating score of user $i$ to item $j$ predicted  from factor $k$ is calculated as:
\begin{equation}
\vspace{-0.1cm}
    \hat{r}_{i,j}^{k}=\mathbf{w}^{\top} \mathbf{h}_{i,j}^{k}
\end{equation}
where $\mathbf{w}\in \mathbb{R}^\frac{d}{K}$ is a parameter vector. 
Then we employ
an attention network over interaction features to
identify the decisive factors and
make the final prediction. The attention weight of
the $k$-th factor $\alpha_k$ is computed as follows:
\begin{equation}
\vspace{-0.1cm}
    \begin{gathered}
a_{k}=\sigma \left(\mathbf{w}_{r}^{\top}  \mathbf{h}_{i,j}^{k}+b_{r}\right) \\
\alpha_{k}=\frac{\exp \left(a_{k}/\tau\right)}{\sum_{k'=1}^{K} \exp \left(a_{k'}/\tau\right)}
\end{gathered}
\end{equation}
where $\mathbf{w}_{r}\in \mathbb{R}^\frac{d}{K}$ and $b_{r}\in \mathbb{R}$ are parameters. The final predicted rating score $\hat{r}_{i,j}$ is the sum of the scores from different factors weighted by their
attention weights, i.e., $\hat{r}_{i,j}=\sum_{k=1}^K \alpha_k
\hat{r}_{i,j}^{k}$, which can be viewed as voting from different factors, consistent with the decision-making process of human rating.

\vspace{-0.1cm}
\subsection{Disentangled Contrastive Learning Module}
Following the supervised learning framework \cite{chen2018neural}, the parameters in DGL and AI modules can be updated by forcing the predicted ratings to be as close as to the observed ratings. However, the sparse interactions restrict the model's capacity of disentangling the latent factors and accurately modeling the user preferences. To alleviate the sparsity issue, we introduce two factor-wise contrastive learning tasks, i.e., node discrimination and edge discrimination.
\vspace{-0.1cm}
\subsubsection{Factor-wise Node Discrimination.} Following \cite{wu2021self}, we perform edge dropping  to generate different views for node discrimination. Specifically, we randomly discard edges with probability $p$ and derive two subgraphs $G_1$ and $G_2$. For user $i$, we employ the DGL module to learn the factorized embeddings $\{\mathbf{u}_{1,i}^k\}_{k=1}^K$ and $\{\mathbf{u}_{2,i}^k\}_{k=1}^K$ on $G_1$ and $G_2$. Then the  discriminative subtask under factor $k$ is to maximize the consistency between the positive pair $(\mathbf{u}_{1,i}^k,\mathbf{u}_{2,i}^k)$ compared with negative pairs $(\mathbf{u}_{1,i}^k,\mathbf{u}_{2,i'}^k)$ where $i\neq i'$. The contrastive loss for user nodes is represented as the expectation of
$K$ subtasks under the latent factors:
\begin{equation}
\begin{aligned}
    \vspace{-0.1cm}
    \mathcal{L}_{fn d}^{u s e r}=&-\mathbb{E}_{\mathcal{K}\times\mathcal{U} }\left[\log  D\left(\mathbf{u}_{1,i}^{k}, \mathbf{u}_{2,i}^{k}\right)\right]+\mathbb{E}_{\mathcal{K}\times\mathcal{U} \times \mathcal{U}^{\prime}  }\left[\log D\left(\mathbf{u}_{1,i}^{k}, \mathbf{u}_{2,i^{\prime}}^{k}\right)\right]
\end{aligned}
\end{equation}
where $i$ is the input user, $i^{\prime}$ is uniformly sampled from $\mathcal{U}^{\prime}=\mathcal{U}$ and $k$ is sampled from $\mathcal{K}$. $D\left(\mathbf{a}, \mathbf{b}\right)=\sigma\left(\mathbf{a}^{\top} \mathbf{W} \mathbf{b}\right)$ is the nonlinear similarity function with trainable parameter $\mathbf{W}$.
Analogously, we can obtain the contrastive loss on item nodes
$\mathcal{L}_{fnd}^{item}$. 
Combining the two losses, we get the
objective function of FND task as $\mathcal{L}_{fnd}=\mathcal{L}_{fnd}^{u s e r}+\mathcal{L}_{fnd}^{item}$.

By factorizing the instance discrimination into K factor-level subtasks, FND can ensure that each disentangled factor of the vectorized representations is sufficiently discriminative. Thus the user/item representations are encouraged to be disentangled and better model the aspect pertinent to one latent factor of users/items.
\vspace{-0.2cm}
\subsubsection{Factor-wise Edge Discrimination.}
To enrich the semantic information in the interaction features, we
further devise a CL paradigm between the factorized interaction features and factorized review features.
 Specifically, for the interaction feature $\mathbf{h}_{i,j}^k$ derived from Equation (8), we treat the review feature $\mathbf{e}_{i,j}^k$ pertinent to factor $k$ as positive sample and other review features $\mathbf{e}_{i^{\prime},j^{\prime}}^k$ as negative samples, where $(i, j)\neq (i^{\prime},j^{\prime})$. Then we make the positive samples closer and negative samples far from each other in the representation
space. And the contrastive loss of FED is represented as:
\begin{equation}
    \mathcal{L}_{fed}=-\mathbb{E}_{\mathcal{K}\times\mathcal{E} }\left[\log D\left(\mathbf{h}_{i,j}^k, \mathbf{e}_{i, j}^k\right)\right]+\mathbb{E}_{\mathcal{K}\times\mathcal{E} \times \mathcal{E}^{\prime}  }\left[\log D\left(\mathbf{h}_{i,j}^k, \mathbf{e}_{i^{\prime},j^{\prime}}^k\right)\right]
\end{equation}
where $(i,j)$ is the input user-item pair and $(i^{\prime},j^{\prime})$ is uniformly sampled from $\mathcal{E}^{\prime}=\mathcal{E}$.
The interpretation here is that we encourage the factorized interaction features to be aware of not only the rating score, but also the reason for user's rating based on factor $k$.
\begin{table}
\centering
\caption{ Statistics of datasets.}
\vspace{-0.2cm}
\begin{tabular}{p{1cm}p{1cm}<{\centering}p{1cm}<{\centering}p{1cm}<{\centering}p{1cm}<{\centering}p{1cm}<{\centering}}
\hline
 \small{\textbf{Datasets}} & \small{Toys} & \small{Clothing} &  \small{Office} & \small{Kitchen} & \small{Tools}\\ 
  \hline
    \hline
 \small{\textbf{\#Users}} &  \small{$19,412$} & 
 \small{$4,905$} & \small{$39,387$} & \small{$66,519$} & \small{$16,638$}  \\
   \small{\textbf{\#Items}}& \small{$11,924$} & \small{$2,420$} &  \small{$23,033$} & \small{$28,237$} & \small{$10,217$}  \\
     \small{\textbf{\#Reviews}} & \small{$167,597$} &
     \small{$53,228$} & \small{$278,677$} & \small{$551,682$} & \small{$134,476$} \\
     \small{\textbf{Density}} & \small{$0.072\%$} &
     \small{$0.448\%$} & \small{$0.031\%$}&
     \small{$0.029\%$} & \small{$0.079\%$}\\
\hline
\hline
\end{tabular}
\label{tab1}
\vspace{-0.5cm}
\end{table}

\begin{table*}
\centering
\caption{Comparison results on the six datasets in terms of MSE. The best and second-best results are highlighted with boldface and underlined. All the
results are reported as the mean value across 5 random runs.}
\vspace{-0.2cm}
\begin{tabular}{p{1cm}p{0.8cm}<{\centering}p{0.8cm}<{\centering}p{1.2cm}<{\centering}p{1.1cm}<{\centering}p{0.9cm}<{\centering}p{0.8cm}<{\centering}p{0.8cm}<{\centering}p{0.8cm}<{\centering}p{0.8cm}<{\centering}p{0.8cm}<{\centering}p{0.8cm}<{\centering}p{0.8cm}<{\centering}p{1.0cm}<{\centering}}
\hline
 \small{\textbf{Datasets}} & \small{\textbf{SVD}} & \small{\textbf{NCF}} &  \small{\textbf{DeepCoNN}} & \small{\textbf{TransNet}} & \small{\textbf{DRRNN}} & \small{\textbf{NARRE}} & \small{\textbf{DAML}}& \small{\textbf{DGCF}} & \small{\textbf{RMG}}& \small{\textbf{RG}} & \small{\textbf{RGCL}} & \small{\textbf{DGCLR}} & \small{\textbf{Improv.}}\\ 
  \hline
    \hline
 \small{\textbf{Toys}} &  \normalsize{$0.8082$} & 
 \normalsize{$0.8075$} & \normalsize{$0.8026$} & \normalsize{$0.7982$} & \normalsize{$0.7884$} & \normalsize{$0.7961$} & \normalsize{$0.7940$}& \normalsize{$0.7943$} & \normalsize{$0.7901$} & \normalsize{$0.7853$}& \normalsize{\underline{$0.7771$}} & \normalsize{\bm{$0.7717$}} & \normalsize{\bm{$0.7\%$}}\\
 \hline
  \small{\textbf{Clothing}} &  \normalsize{$1.1161$} & 
 \normalsize{$1.1094$} & \normalsize{$1.1184$} & \normalsize{$1.1141$} & \normalsize{$1.1035$} & \normalsize{$1.1064$} & \normalsize{$1.1065$}& \normalsize{$1.1002$} & \normalsize{$1.1064$} & \normalsize{$1.1024$}& \normalsize{\underline{$1.0858$}} & \normalsize{\bm{$1.0573$}} & \normalsize{\bm{$2.6\%$}}\\
 \hline
     \small{\textbf{Office}}&  \normalsize{$0.7438$} & 
 \normalsize{$0.7459$} & \normalsize{$0.7426$} & \normalsize{$0.7419$} & \normalsize{$0.7306$} & \normalsize{$0.7408$} & \normalsize{$0.7358$}& \normalsize{$0.7345$} & \normalsize{$0.7348$} & \normalsize{$0.7293$}& \normalsize{\underline{$0.7228$}} & \normalsize{\bm{$0.7146$}} & \normalsize{\bm{$1.1\%$}}\\
 \hline
     \small{\textbf{Kitchen}} &  \normalsize{$1.1011$} & 
 \normalsize{$1.0946$} & \normalsize{$1.0914$} & \normalsize{$1.0879$} & \normalsize{$1.0769$} & \normalsize{$1.0835$} & \normalsize{$1.0814$}& \normalsize{$1.0798$} & \normalsize{$1.0783$} & \normalsize{$1.0754$}& \normalsize{\underline{$1.0732$}} & \normalsize{\bm{$1.0658$}} & \normalsize{\bm{$0.7\%$}} \\
 \hline
     \small{\textbf{Tools}}&  \normalsize{$0.9412$} & 
 \normalsize{$0.9385$} & \normalsize{$0.9356$} & \normalsize{$0.9348$} & \normalsize{$0.9249$} & \normalsize{$0.9304$} & \normalsize{$0.9295$}& \normalsize{$0.9301$} & \normalsize{$0.9288$} & \normalsize{$0.9253$}& \normalsize{\underline{$0.9241$}} & \normalsize{\bm{$0.9145$}} & \normalsize{\bm{$1.0\%$}} \\
\hline
\hline
\end{tabular}
\label{tab2}
\vspace{-0.4cm}
\end{table*}

\vspace{-0.1cm}
\subsection{Model Optimization}
Following previous rating prediction works \cite{zheng2017joint,mnih2007probabilistic}, we employ Mean Square Error (MSE) loss as the supervision signal:
\begin{equation}
    \mathcal{L}_{\text {sup}}=\frac{1}{|\mathcal{T}|} \sum_{(i, j) \in \mathcal{T}}\left(\hat{r}_{i j}-r_{i j}\right)^{2},
\end{equation}
where $\mathcal{T}$ denotes the observed user-item interactions in the training set. After combining the above two factor-wise contrastive losses, the overall  optimization target is represented as:
\begin{equation}
    \mathcal{L}=\mathcal{L}_{\text {sup}}+\lambda_1 \mathcal{L}_{fnd}+\lambda_2 \mathcal{L}_{fed}
\end{equation}
where $\lambda_1$ and $\lambda_2$ are hyperparameters to control
the contributions of CL tasks towards the overall objective.
\vspace{-0.3cm}
\subsection{Model Complexity Analysis}
For the memory cost, it is notable that we divide the ID embedding into $K$ chunks to keep that the same as previous works \cite{liu2021learning,xi2021deep}. The extra parameters involved in the DGL, AI and DCL modules are $O(d \times d)$, $O(\frac{d}{K} \times \frac{d}{K})$ and $O(\frac{d}{K} \times \frac{d}{K})$, respectively. For the time cost, the complexity of the DGL and AI module is $O(L\times K \times |\mathcal{E}| \times \frac{d}{K})$ and $O(K \times(M+N+|\mathcal{E}|)\times \frac{d}{K})$. In the DCL module, the complexity of FND and FED is $O(K\times(M+N)\times B \times \frac{d}{K} )$ and $O(K\times |\mathcal{E}|\times B \times \frac{d}{K})$, respectively. Here $B$ is the number of negative samples for each positive sample. Since we set $B=1$ in practice, the overall time complexity of DGCLR is $O(L\times |\mathcal{E}| \times d)$, which is the same with GNN-based recommendation methods \cite{shuai2022review,xia2022hypergraph}.
\vspace{-0.1cm}

\section{Experiments}
Our experiments aim to answer the following research questions:
\begin{itemize}
    \item \textbf{RQ1}: How does DGCLR compare to state-of-the-art methods in rating prediction tasks?
    \item \textbf{RQ2}: How do the proposed DGL, AI, and DCL modules contribute to DGCLR's performance?
    \item \textbf{RQ3}: How do the key hyperparameters influence DGCLR's performance?
    \item \textbf{RQ4}: Can DGCLR offer interpretability of the learned factor graphs and rating prediction?
\end{itemize}
\vspace{-0.3cm}
\subsection{Experimental Settings}
\subsubsection{Datasets}
Following previous works \cite{xi2021deep,shuai2022review}, we  evaluate DGCLR on the Amazon review
dataset \cite{he2016ups}. \textit{Toys and Games}, \textit{Office Products}, \textit{Clothing}, \textit{Home and Kitchen}, and \textit{Tools and Home Improvement} are the five 5-core subsets selected (shortened as Toys, Office, Clothing, Kitchen, and Tools, respectively). The
rating scores for all the five datasets range from 1 to 5. Each dataset is randomly split into training, validation, and testing sets with a ratio of 8:1:1. The details of these datasets are summarized in Table \ref{tab1}.
\vspace{-0.1cm}
\subsubsection{Baselines.}
We compare DGCLR with state-of-the-art
methods, including traditional rating-based CF methods (SVD and NCF), CNN-based methods (DeepCoNN, TransNet, and DRRNN), attention-based methods (NARRE and DAML), disentanglement-based method (DGCF), and graph-based methods (RMG, RG, and RGCL):
\begin{itemize}
    \item \textbf{SVD} \cite{koren2009matrix} is a matrix factorization model that uses the inner product of the latent factors of users and items to estimate ratings.
    \item \textbf{NCF} \cite{he2017neural} replaces the inner product with a neural network to predict the rating based on user and item free embeddings.
    \item \textbf{DeepCoNN} \cite{zheng2017joint} models user behaviors and item properties from review data using  two parallel networks.
    \item \textbf{TransNet} \cite{catherine2017transnets} extends DeepCoNN  by adding an additional layer for learning the target review features.
    \item \textbf{DRRNN} \cite{xi2021deep} uses both target ratings and reviews for backpropagation to retain more semantic review information.
    \item \textbf{NARRE} \cite{chen2018neural} employs an attention mechanism to model reviews and a neural regression model for rating prediction.
    \item \textbf{DAML} \cite{liu2019daml} uses the local and mutual attention of CNN to learn the user-item interaction and user/item representations.
    \item \textbf{DGCF} \cite{wang2020disentangled} is a collaborative filtering method that employs a neighbor routing mechanism to disentangle the user-item graph for fine-grained user/item representation learning.
    \item \textbf{RMG} \cite{wu2019reviews} learns user/item representations from both review-content and graph views via  a three-level attention network.
    \item \textbf{RG} \cite{shuai2022review} learns user/item representations based on user-item graph with review feature-enhanced edges.
    \item \textbf{RGCL} \cite{shuai2022review} is the SOTA method that is further equipped with two contrastive learning modules based on RG.
    \vspace{-0.1cm}
\end{itemize}
It is notable that we reimplement DGCF by replacing the BPR loss \cite{rendle2012bpr} with MSE loss to accommodate the rating prediction task.
\begin{figure}\setlength{\abovecaptionskip}{-0.05cm}
\centering
\subfigure[Clothing]{
\includegraphics[width=4.1cm]{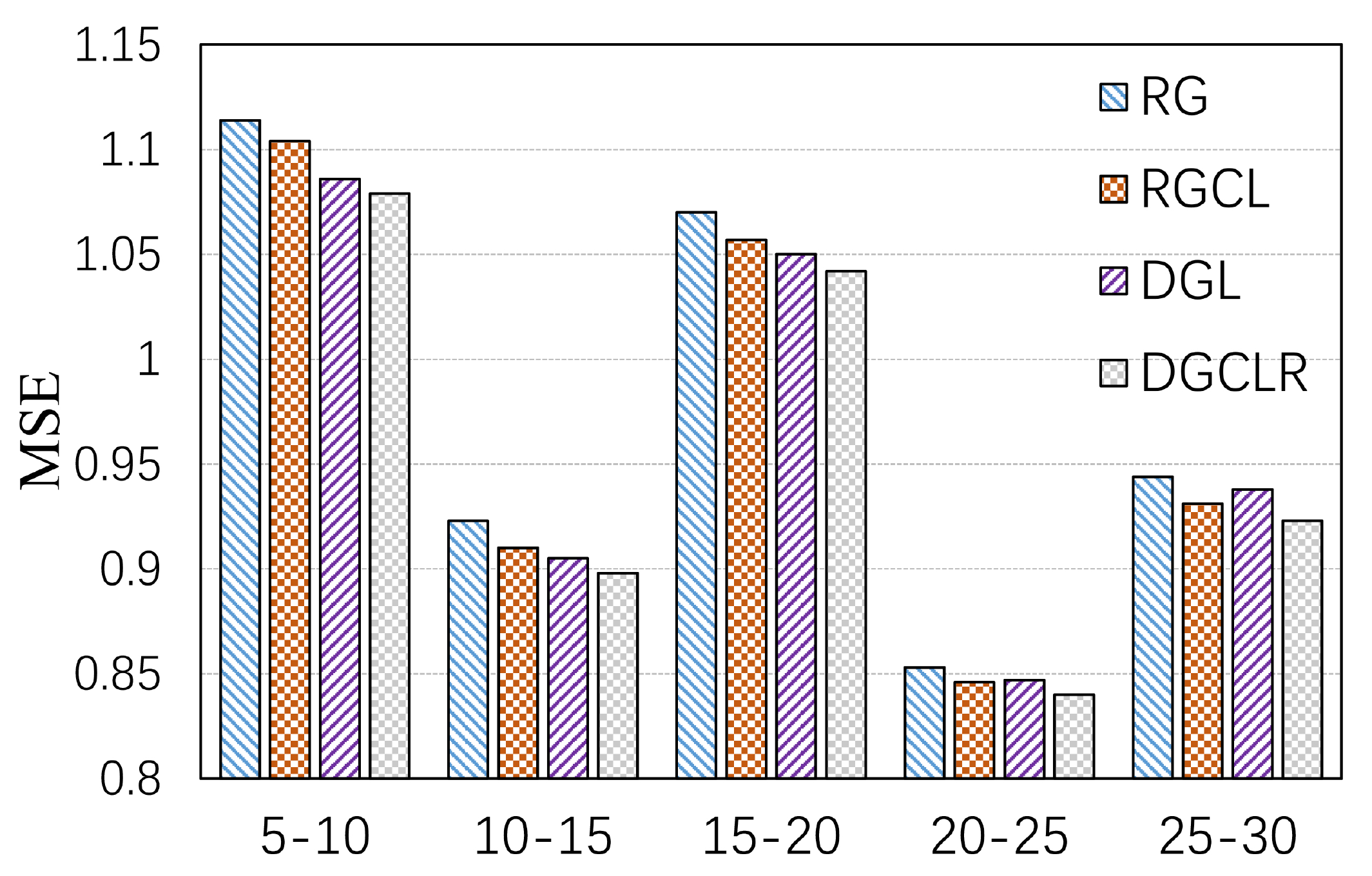}
}
\quad
\hspace{-.26in}
\subfigure[Office]{
\includegraphics[width=4.1cm]{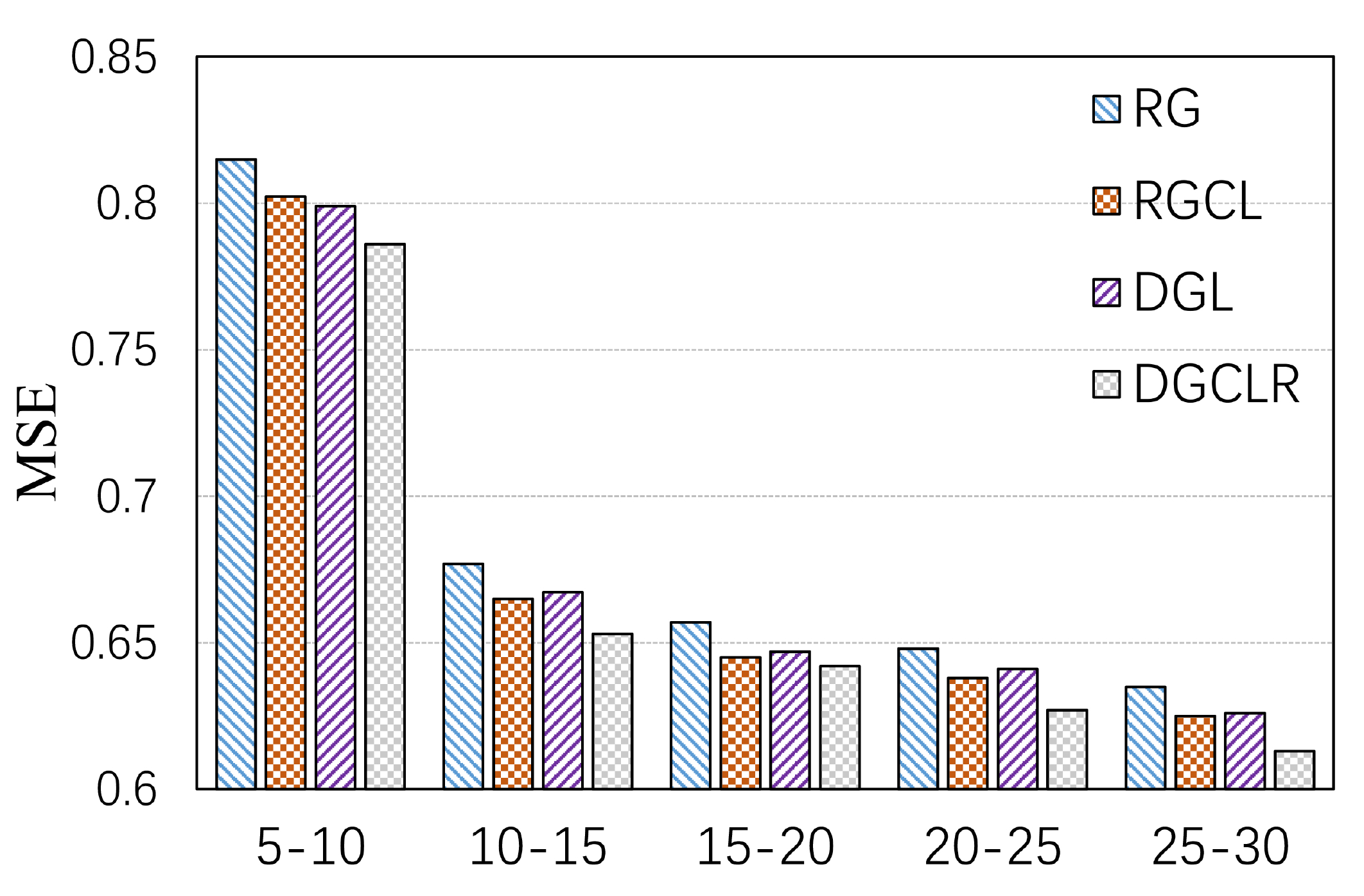}
}
\caption{Performance \textit{w.r.t} interaction degrees.}
\label{fig6}
\vspace{-0.4cm}
\end{figure}
\vspace{-0.1cm}
\subsubsection{Evaluation Metric.}
Following \cite{xi2021deep,shuai2022review}, we evaluate the performance by MSE. Each experiment is repeated five times. We report the average accuracy for each dataset. As suggested by previous works \cite{li2021recommend,tay2018multi}, a relative improvement of more than 1\% is considered significant.
\vspace{-0.3cm}
\subsubsection{Parameter Settings}
The hyperparameters for the baseline models are tuned according to the original paper. For DGCLR, we use the Xavier method \cite{glorot2010understanding} to initialize all trainable parameters and Adam \cite{kingma2014adam} to optimize the parameters with a learning rate of 0.01.
Following \cite{liu2019daml,xi2021deep}, the size of embeddings $d$ for users/items and reviews is chosen from \{32, 64, 128\}. We choose the number of message passing layers $L$ from \{1, 2, 3\}, the number of latent factors from \{2, 4, 8\}, and the dropout ratio from \{0.7, 0.8, 0.9\}. The temperature hyperparameter $\tau$ is selected from \{0.2, 0.5, 1\}.
The hyperparameters $\lambda_1$ and $\lambda_2$ are searched from \{0.1, 0.3, 0.5, 0.7, 0.9\}. 
\vspace{-0.3cm}
\subsection{Performance Comparison (RQ1)}
\subsubsection{Overall performance comparison.}
The comparison results of all methods are presented in Table \ref{tab2}. Based on the results, the following observations can be made:
\begin{table}
\centering
\caption{Ablation studies on the DGL and AI modules.}
\vspace{-0.2cm}
\begin{tabular}{p{1.5cm}p{1.5cm}<{\centering}p{1.5cm}<{\centering}p{1.3cm}<{\centering}}
\hline
 \small{\textbf{Datasets}} & \normalsize{Toys} & \normalsize{Clothing} &  \normalsize{Office}\\ 
  \hline
    \hline
 \normalsize{\textbf{RG}} &  \normalsize{$0.7853$} & 
 \normalsize{$1.1024$} & \normalsize{$0.7293$} \\
 \hline
   \normalsize{\textbf{Variant 1}}&  \normalsize{$0.7857$} & 
 \normalsize{$1.0987$} & \normalsize{$0.7287$} \\
     \normalsize{\textbf{Variant 2}} &  \normalsize{$0.7801$} & 
 \normalsize{$1.0793$} & \normalsize{$0.7231$} \\
     \normalsize{\textbf{Variant 3}} &  \normalsize{$0.7841$} & 
 \normalsize{$1.0871$} & \normalsize{$0.7266$} \\
 \hline
 \normalsize{\textbf{DGL}} &  \normalsize{$0.7793$} & 
 \normalsize{$1.0728$} & \normalsize{$0.7222$} \\
  \hline
  \normalsize{\textbf{DGL+AI}} &  \normalsize{\bm{$0.7776$}} & 
 \normalsize{\bm{$1.0691$}} & \normalsize{\bm{$0.7210$}} \\
\hline
\hline
\end{tabular}
\label{tab3}
\vspace{-0.2cm}
\end{table}
\begin{table}
\centering
\caption{Ablation studies on the DCL module.}
\vspace{-0.2cm}
\begin{tabular}{p{1.9cm}p{1.5cm}<{\centering}p{1.5cm}<{\centering}p{1.3cm}<{\centering}}
\hline
 \small{\textbf{Datasets}} & \normalsize{Toys} & \normalsize{Clothing} &  \normalsize{Office}\\ 
  \hline
    \hline
 \normalsize{\textbf{DGL+AI+ND}} &  \normalsize{$0.7762$} & 
 \normalsize{$1.0634$} & \normalsize{$0.7188$} \\
   \normalsize{\textbf{DGL+AI+FND}}&  \normalsize{$0.7743$} & 
 \normalsize{$1.0586$} & \normalsize{$0.7165$} \\
  \hline
     \normalsize{\textbf{DGL+AI+ED}} &  \normalsize{$0.7755$} & 
 \normalsize{$1.0641$} & \normalsize{$0.7184$} \\
 \normalsize{\textbf{DGL+AI+FED}} &  \normalsize{$0.7732$} & 
 \normalsize{$1.0588$} & \normalsize{$0.7173$} \\
  \hline
  \normalsize{\textbf{DGCLR}} &  \normalsize{\bm{$0.7717$}} & 
 \normalsize{\bm{$1.0573$}} & \normalsize{\bm{$0.7146$}} \\
\hline
\hline
\end{tabular}
\label{tab4}
\vspace{-0.2cm}
\end{table}
\begin{itemize}
\vspace{-0.1cm}
    \item DGCLR achieves the best results on every dataset tested and significantly outperforms the strongest baseline, RGCL, on three out of five datasets. The improvements of DGCLR relative to all other baselines can be attributed to: 1) By disentangling the graph from semantic and structural perspectives, DGCLR is able to model user preferences based on multiple latent factors more accurately. 2) The AI module enables DGCLR to make rating predictions by taking into account all latent interaction factors. 3) The FND and FED tasks can assist DGCLR in disentangling the factors and incorporating review data more effectively into learned user/item and interaction representations.
    \item In general, the performances of GNN-based methods (DGCF, RMG, RG, RGCL) are superior to other methods. This phenomenon demonstrates the effectiveness of graph learning in processing higher-order information from multi-hop neighborhoods.
    \item DGCF achieves comparable or superior performance in comparison to many CNN-based or attention-based baselines despite its ignoring review information. This also demonstrates the efficacy of disentangling and graph learning in review-based recommender systems. Meanwhile, DGCLR has an average improvement of 2\% compared to DGCF, which validates the efficacy of the graph disentangling mechanism in DGCLR and highlights the importance of review data in disentangling.
    \vspace{-0.1cm}
\end{itemize}
\vspace{-0.1cm}
\subsubsection{Performance comparison in alleviating data sparsity.} To verify the robustness of DGCLR against sparsity issue, we partition users into distinct groups according to their interaction numbers in the training set (e.g., 5-10). Then we report the MSE of DGL (DGCLR minus the DCL module), DGCLR in comparison to the SOTA models RG and RGCL for each group. Figure \ref{fig6} demonstrates that, compared to RG, DGL is more robust to the sparsity issue, allowing for more effective use of review information to disentangle latent factors in user-item interactions. In addition, DGCLR improves upon RGCL by conducting contrastive learning in each representation subspace of a factor independently, rather than in
the whole representation space. Consequently, DGCLR
achieves the highest performance across all groups, demonstrating that our proposed factor-wise CL tasks can better alleviate the sparsity issue than the entangled CL tasks in RGCL.

\begin{table}
\centering
\caption{Impact of latent factor number on DGCLR.}
\vspace{-0.2cm}
\begin{tabular}{p{1.3cm}p{1.5cm}<{\centering}p{1.5cm}<{\centering}p{1.3cm}<{\centering}}
\hline
 \small{\textbf{Datasets}} & \normalsize{Toys} & \normalsize{Clothing} &  \normalsize{Office}\\ 
  \hline
    \hline
 \normalsize{\textbf{$K=1$}} &  \normalsize{$0.7797$} & 
 \normalsize{$1.0817$} & \normalsize{$0.7231$} \\
   \normalsize{\textbf{$K=2$}}&  \normalsize{$0.7723$} & 
 \normalsize{\bm{$1.0575$}} & \normalsize{$0.7164$} \\
     \normalsize{\textbf{$K=4$}} &  \normalsize{\bm{$0.7717$}} & 
 \normalsize{$1.0595$} & \normalsize{\bm{$0.7146$}} \\
 \normalsize{\textbf{$K=8$}} &  \normalsize{$0.7772$} & 
 \normalsize{$1.0639$} & \normalsize{$0.7217$} \\
\hline
\hline
\end{tabular}
\label{tab5}
\vspace{-0.6cm}
\end{table}
\begin{figure}\setlength{\abovecaptionskip}{-0.1cm}
\centering
\subfigure[Toys]{
\includegraphics[width=2.78cm]{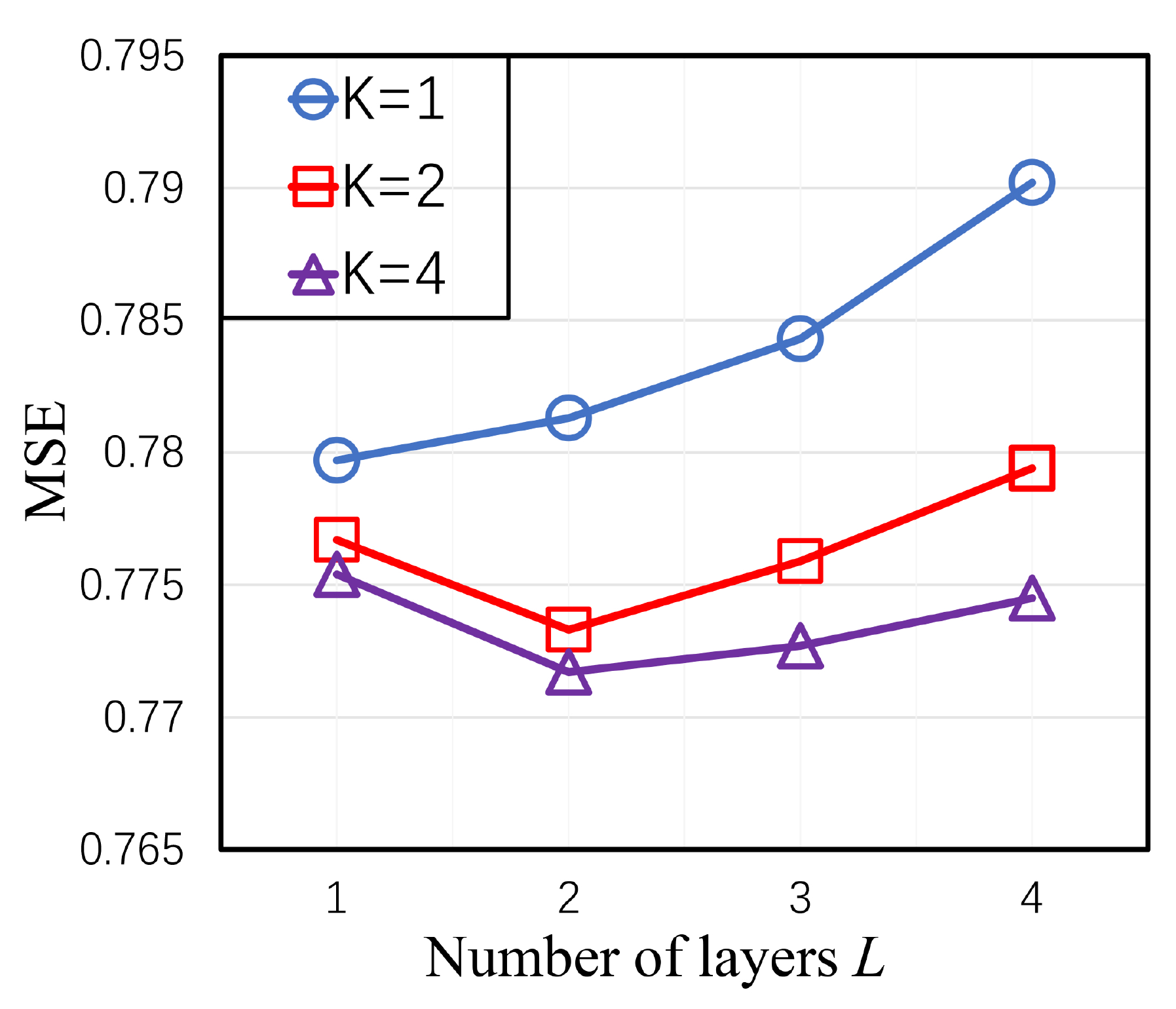}
}
\quad
\hspace{-.26in}
\subfigure[Clothing]{
\includegraphics[width=2.78cm]{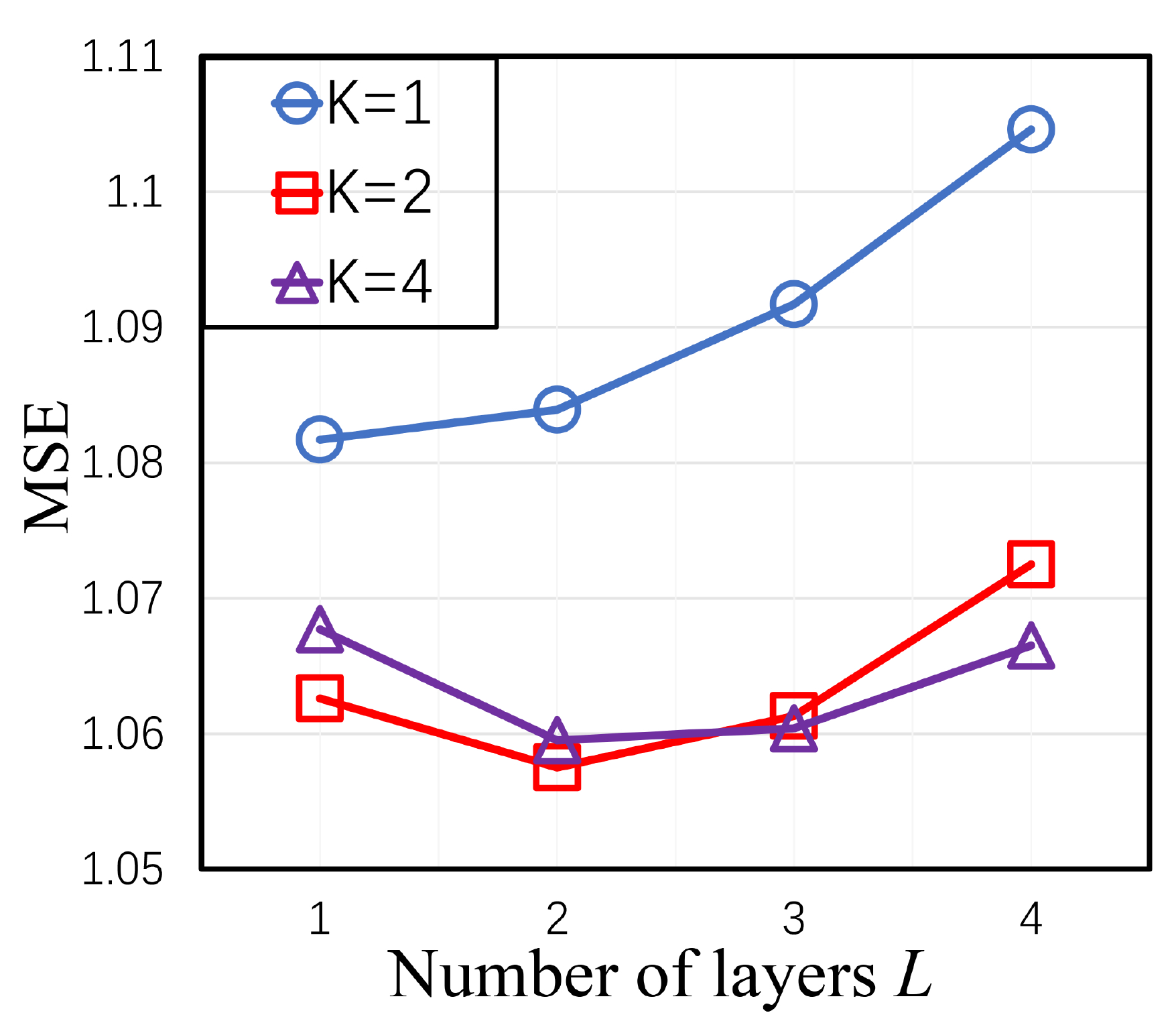}
}
\quad
\hspace{-.26in}
\subfigure[Office]{
\includegraphics[width=2.78cm]{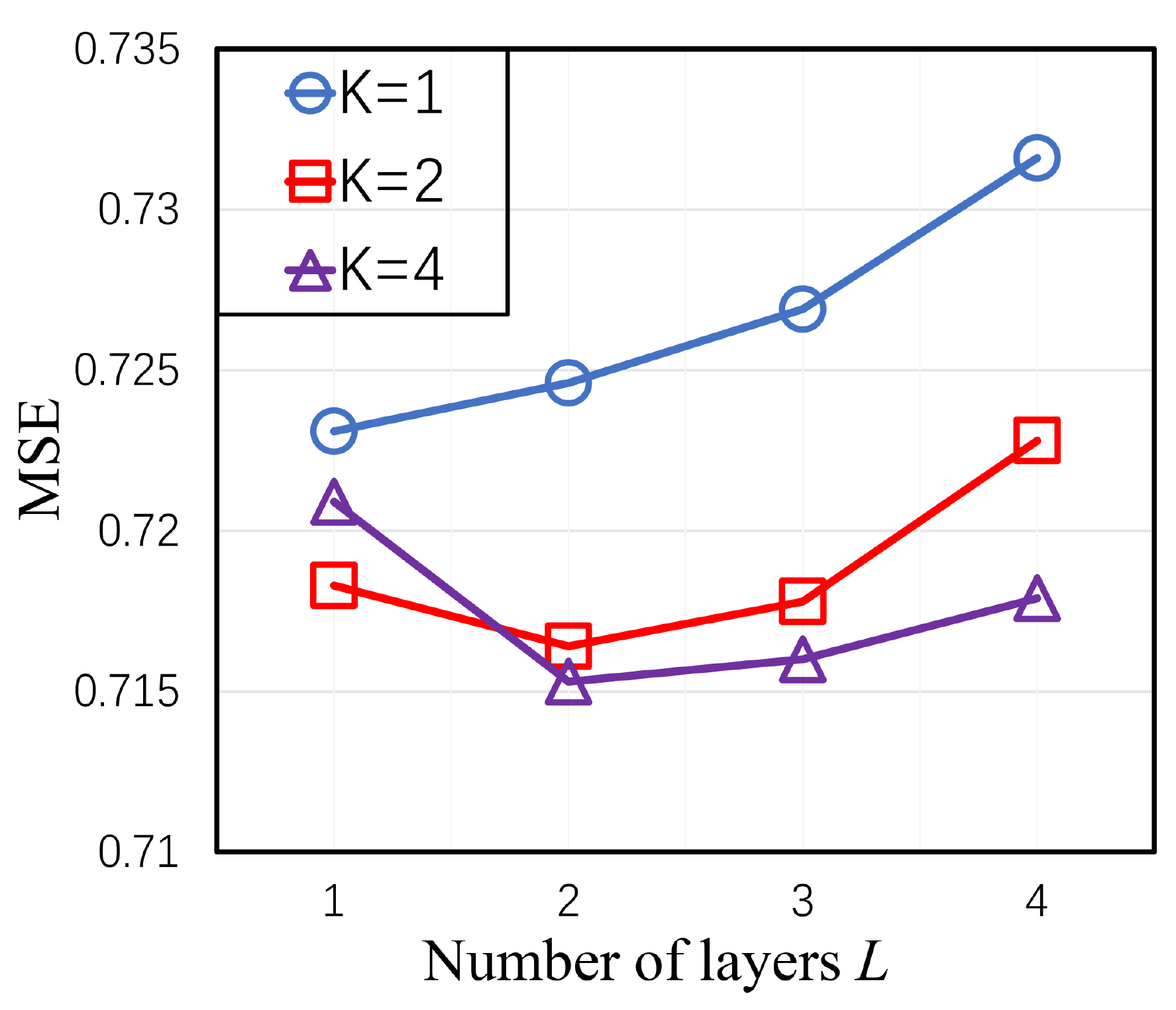}
}
\caption{Impact of layer number $L$ on DGCLR.}
\label{fig3}
\vspace{-0.7cm}
\end{figure}
 \begin{table*}
\centering
\caption{Examples of reviews corresponding to each latent factor on Office. The key information is highlighted with red.}
\vspace{-0.2cm}
\begin{tabular}{p{1.1cm}p{0.7cm}<{\centering}p{1.6cm}<{\centering}p{7cm}<{\centering}}
\hline
 \multicolumn{1}{|c|}{\multirow{3}{1.1cm}{\small{Factor $k_1$}}} & \multicolumn{1}{c}{\multirow{1}{0.7cm}{\small{$r=1$}}} & \multicolumn{1}{|c|}{\small{$s_{i,j}^{k_1,L}=0.547$}} &  \multicolumn{1}{l|}{\small{\text{Although i love the pastel colors, this item is \textcolor{red}{wasteful}. Unfortunately, I 'll \textcolor{red}{never use the note tabs}.}}}\\ 
 \cline{2-4}
  \multicolumn{1}{|c|}{} & \multicolumn{1}{c}{\multirow{1}{0.7cm}{\small{$r=3$}}}  & \multicolumn{1}{|c|}{\small{$s_{i,j}^{k_1,L}=0.538$}} &  \multicolumn{1}{l|}{\small{\text{This product is okay, but i had a difficult time getting it to stay open. i \textcolor{red}{don't think} it would be very \textcolor{red}{beneficial in my business}.}}}\\ 
   \cline{2-4}
  \multicolumn{1}{|c|}{} & \multirow{1}{0.7cm}{\small{$r=5$}}& \multicolumn{1}{|c|}{\small{$s_{i,j}^{k_1,L}=0.615$}} & \multicolumn{1}{l|}{\small{the range is good and the clarity can not be beat in my opinion. the options are just \textcolor{red}{what i needed for my purposes}.}}\\
   \hline
   
 \multicolumn{1}{|c|}{\multirow{3}{1.1cm}{\small{Factor $k_2$}}} & \multicolumn{1}{c}{\multirow{1}{0.7cm}{\small{$r=1$}}} & \multicolumn{1}{|c|}{\small{$s_{i,j}^{k_2,L}=0.551$}} &  \multicolumn{1}{l|}{\small{\text{I had previously given this a five star review, but after two months the stapler \textcolor{red}{jammed shut} and \textcolor{red}{would not open}.}}}\\ 
 \cline{2-4}
  \multicolumn{1}{|c|}{} & \multicolumn{1}{c}{\multirow{1}{0.7cm}{\small{$r=3$}}}  & \multicolumn{1}{|c|}{\small{$s_{i,j}^{k_2,L}=0.587$}} &  \multicolumn{1}{l|}{\tabincell{l}{\small{\text{Print \textcolor{red}{quality seems to be OK}, but there 's no way to tell the printer whether you are using plain paper of glossy paper. Plain}}\\ \small{\text{paper prints look washed out, premium paper prints look good.
}}}}\\ 
   \cline{2-4}
  \multicolumn{1}{|c|}{} & \multirow{1}{0.7cm}{\small{$r=5$}}& \multicolumn{1}{|c|}{\small{$s_{i,j}^{k_2,L}=0.513$}} & \multicolumn{1}{l|}{\small{It's \textcolor{red}{solidly made} and \textcolor{red}{stands up to regular use} pretty darn well. The result is crisp laser printing on a home office budget.}}\\
   \hline
   
    \multicolumn{1}{|c|}{\multirow{3}{1.1cm}{\small{Factor $k_3$}}} & \multicolumn{1}{c}{\multirow{1}{0.7cm}{\small{$r=1$}}} & \multicolumn{1}{|c|}{\small{$s_{i,j}^{k_3,L}=0.637$}} &  \multicolumn{1}{l|}{\small{\text{My rating reflects my dissatisfaction with this vendors \textcolor{red}{deceptive advertising}.}}}\\ 
 \cline{2-4}
  \multicolumn{1}{|c|}{} & \multicolumn{1}{c}{\multirow{1}{0.7cm}{\small{$r=3$}}}  & \multicolumn{1}{|c|}{\small{$s_{i,j}^{k_3,L}=0.526$}} &  \multicolumn{1}{l|}{\small{\text{This product almost \textcolor{red}{delivers on its promises one}. But the individual packets of labels easily detached from the main package.
}}}\\ 
   \cline{2-4}
  \multicolumn{1}{|c|}{} & \multirow{1}{0.7cm}{\small{$r=5$}}& \multicolumn{1}{|c|}{\small{$s_{i,j}^{k_3,L}=0.581$}} & \multicolumn{1}{l|}{\small{Making photo prints uses a lot of ink. This helps address that problem. Same \textcolor{red}{quality} prints as \textcolor{red}{standard capacity cartridge}.}}\\
   \hline
    \multicolumn{1}{|c|}{\multirow{3}{1.1cm}{\small{Factor $k_4$}}} & \multicolumn{1}{c}{\multirow{1}{0.7cm}{\small{$r=1$}}} & \multicolumn{1}{|c|}{\small{$s_{i,j}^{k_4,L}=0.579$}} &  \multicolumn{1}{l|}{\tabincell{l}{\small{\text{The ink is very \textcolor{red}{inexpensive} but with the \textcolor{red}{quality} of the system so \textcolor{red}{cheaply} made the \textcolor{red}{inexpensive} ink is hardly \textcolor{red}{worth} the cost}} \\ \small{\text{of having to buy a new system in less than two years.}}}}\\ 
 \cline{2-4}
  \multicolumn{1}{|c|}{} & \multicolumn{1}{c}{\multirow{1}{0.7cm}{\small{$r=3$}}}  & \multicolumn{1}{|c|}{\small{$s_{i,j}^{k_4,L}=0.612$}} &  \multicolumn{1}{l|}{\small{\text{For a relatively \textcolor{red}{inexpensive} laminator, this does an OK job. But the \textcolor{red}{lack of guides} on this unit is a real problem.}}}\\ 
   \cline{2-4}
  \multicolumn{1}{|c|}{} & \multirow{1}{0.7cm}{\small{$r=5$}}& \multicolumn{1}{|c|}{\small{$s_{i,j}^{k_4,L}=0.547$}} & \multicolumn{1}{l|}{\small{I bought this \textcolor{red}{because of the price} and to my surprise it is fantastic. I will buy this again over any other more \textcolor{red}{expensive} ones.}}\\
   \hline
\end{tabular}
\label{tab6}
\vspace{-0.2cm}
\end{table*}
\vspace{-0.1cm}
\subsection{Ablation Studies (RQ2)}
In this section, we conduct ablation research on the three modules in DGCLR to comprehend their functions more deeply.
\vspace{-0.2cm}
\subsubsection{Impact of the DGL module.} To validate the efficacy of DGL, we temporarily remove the AI and DCL modules and compare it with RG, the SOTA graph learning model. In particular, we concatenate the factorized user/item representations to obtain their final representations and employ the same interaction module in RG for a fair comparison. In addition to RG, we also compare DGL to its three variants: 1) Variant 1 calculates the coefficient $\mathbf{s}_{i,j}^{k,(l)}$ in a uniform manner, i.e., $\mathbf{s}_{i,j}^{k,(l)}=\frac{1}{K}$. 2) Variant 2 calculates $\mathbf{s}_{i,j}^{k,(l)}$ based on the semantic information, i.e., $\mathbf{s}_{i,j}^{k,(l)}=\mathbf{se}_{i,j}^{k}$. 3) Variant 3 calculates $\mathbf{s}_{i,j}^{k,(l)}$ based on the structural information, i.e., $\mathbf{s}_{i,j}^{k,(l)}=\mathbf{st}_{i,j}^{k,(l)}$. The results are shown in Table \ref{tab3}. The key observations are as follows:

First, the performance of Variant 1 is comparable to that of RG because it fails to model the different distributions among interactions. Second, we observe a decrease in the performance of Variant 2 and 3 compared to DGL, demonstrating that integrating
semantic and structural information allows for a comprehensive exploration of the distributions of latent factors in interactions. Third, the consistently superior performance of Variant 2 over Variant 3 suggests that review information plays a dominant role in identifying the latent factors, which is consistent with the assumption made in Section 4.1.2. Lastly, DGL has a significant improvement over RG, validating the efficacy of disentanglement.

\begin{figure}\setlength{\abovecaptionskip}{-0.1cm}
\centering
\subfigure[Clothing]{
\includegraphics[width=4.1cm]{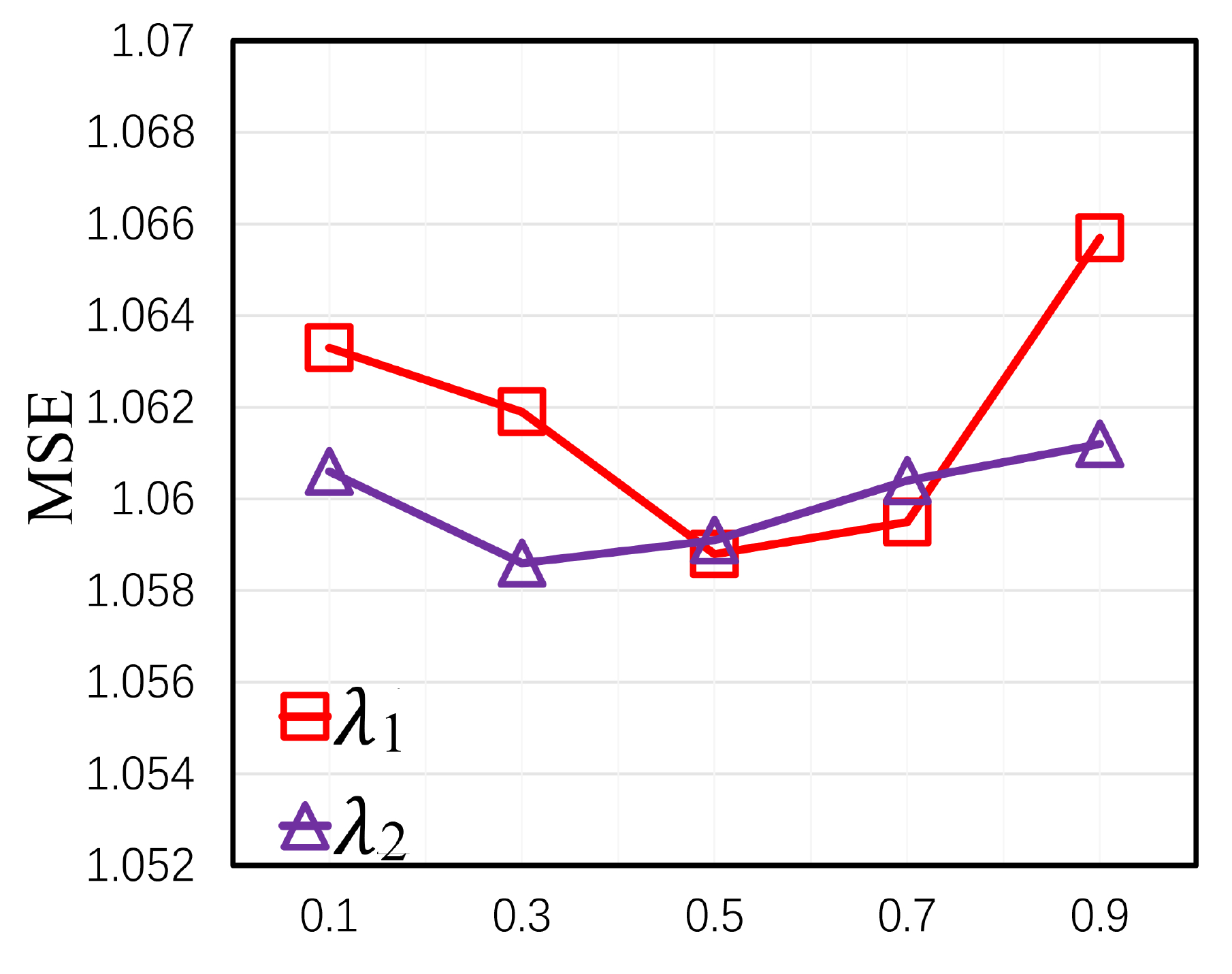}
}
\quad
\hspace{-.26in}
\subfigure[Office]{
\includegraphics[width=4.1cm]{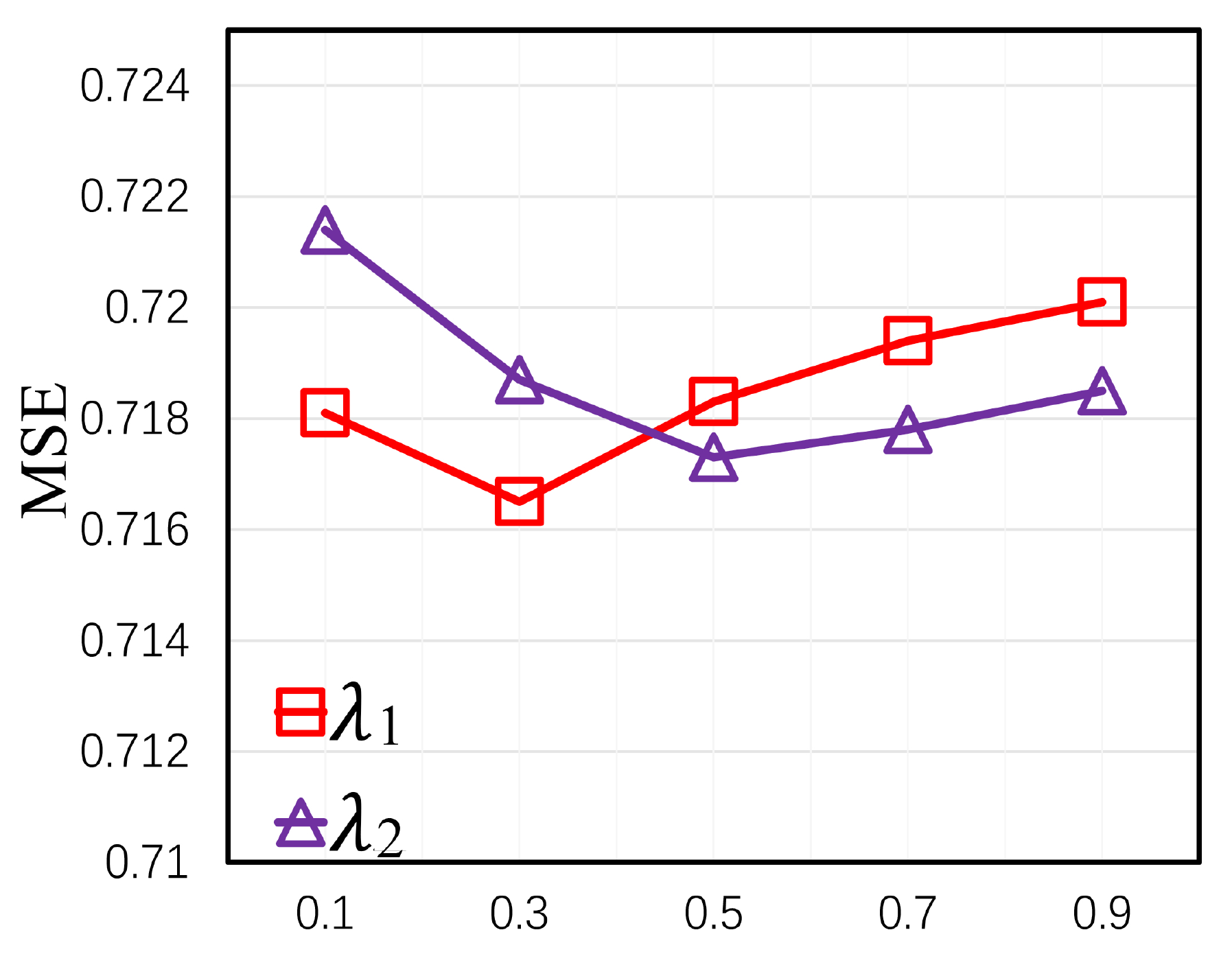}
}
\caption{Impact of $\lambda_1$ and $\lambda_2$ on DGCLR.}
\label{fig4}
\vspace{-0.5cm}
\end{figure}
\vspace{-0.2cm}
\subsubsection{Impact of the AI module.} We further evaluate the impact of AI module by incorporating it into DGL. As shown at the bottom of Table \ref{tab3}, supercharging DGL with the AI module consistently improves performance. This result verifies the capacity of AI module to identify the diverse decisive latent factors when predicting the final rating scores of different users given different items.
\vspace{-0.2cm}
\subsubsection{Impact of the DCL module.}
To validate the efficacy of the two proposed factor-wise CL tasks, FND and FED, we build a variant (ND and ED) for each task by concatenating the learned disentangled representations of user/item and interaction. Then we perform CL tasks on the holistic representations and substitute $e_{i,j}^k$ in Equation (12) with $e_{i,j}$. These two variants degrade to the entangled CL tasks and fail to incorporate factor-level information for disentanglement. Table \ref{tab4} summarizes the results.

We observe that our model with factor-wise CL tasks consistently outperforms that with the holistic CL tasks, highlighting the usefulness of our proposed CL paradigm in facilitating disentanglement. Moreover, by combining the two factor-wise CL tasks, DGCLR achieves the best performance, validating that the factor-wise supervision signals can reinforce the factorized embedding learning through self-discrimination.

\vspace{-0.2cm}
\subsection{Hyperparameter Studies (RQ3)}
 In this section, we evaluate the effect of key hyperparameters (\# of latent factors $K$, \# of message passing layers $L$, and importance hyperparameters of two CL tasks $\lambda_1,\lambda_2$) in DGCLR and show the evaluation results in Table \ref{tab5}, Figure \ref{fig3} and Figure \ref{fig4}.
 \begin{itemize}
 \vspace{-0.1cm}
     \item The value of $K$ is examined in $\{1,2,4,8\}$. According to Table \ref{tab5}, DGCLR performs the worst when $K=1$, indicating that modeling the user/item characteristics as a whole is insufficient to capture user behavioral patterns. Increasing the factor number to $2$ or $4$ significantly enhances performance. However, excessive disentanglement (e.g. $K=8$) leads to performance degradation due to too fine-grained factors and limited expressiveness (e.g. $\frac{d}{K}=8$ when $d=64$ and $K=8$). 
     \item We further investigate the impact of message passing layer number $L$ under varying numbers of factors $K$ by setting $L \in \{1, 2, 3, 4\}$   and $K\in\{1,2,4\}$. As shown in Figure \ref{fig3}, the performance keeps declining with the increasing of $L$ when $K=1$. One possible reason is that multi-layer user, item and review feature propagation brings more entangled and irrelevant information to holistic node representations. This observation aligns with RG \cite{shuai2022review}. Nevertheless, by disentangling the latent factors, DGCLR benefits from higher-order information and achieves a significant performance gain by setting $L$ to a larger value when $K=2$ and $K=4$. In particular, DGCLR achieves the best performance when $L=2$, indicating that second-order connectivity may be adequate for extracting factor-relevant information.
     \item We also study the impact of $\lambda_1, \lambda_2 \in \{0.1,0.3,0.5,0.7,0.9\}$ that weigh the contributions of CL tasks. As shown in Figure \ref{fig4}, the performance of DGCLR with FND or FED both first increases and then decreases. The optimal choices for $\lambda_1$ and  $\lambda_2$ vary among different datasets depending on the data size and sparsity, and fall into the range from 0.3 to 0.7. Generally, DGCLR is not very sensitive when
    $\lambda_1$ and  $\lambda_2$ are tuned in a reasonable range.
    \vspace{-0.1cm}
 \end{itemize}
 \begin{figure}\setlength{\abovecaptionskip}{-0.05cm}
\centering
\centering
\includegraphics[width=3.5in]{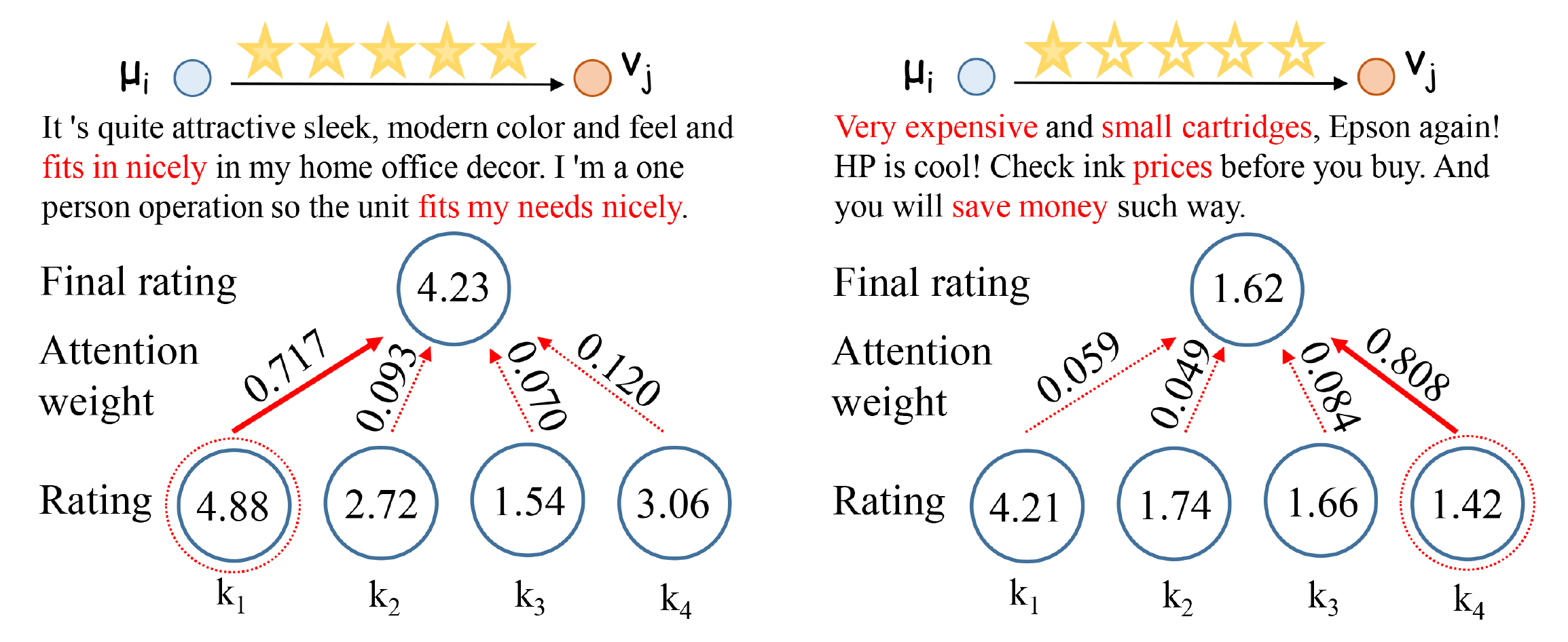}
\centering
\caption{Examples of rating predictions in DGCLR.} 
\label{fig5}
\vspace{-0.4cm}
\end{figure}
\vspace{-0.1cm}
\subsection{Explainability Studies (RQ4)}
\subsubsection{Interpretability of factor graphs.}
To better interpret the latent semantics of the learned factor graphs, we explain the reasons behind user ratings by presenting the reviews of high-confidence interactions. In particular, we conduct experiments on Office with factor number $K=4$ and layer number $L=2$.
For each factor $k$, we randomly select one review with score $\mathbf{s}_{i,j}^{k,L}>0.5$ for rating 1, 3, 5, respectively. This indicates that factor $k$ dominates user $i$'s rating on item $j$. We present the reviews and associated scores in Table \ref{tab6} and have observations as follows:
\begin{itemize}
\vspace{-0.1cm}
    \item By jointly analyzing the reviews of the same latent factor, we find that, despite being written for different items and ratings, they all have inherent semantic connections. For example, the reviews of factor $k_1$ reflect user ratings based on how well the product meets their needs, such as \textit{wasteful} and \textit{beneficial in my business}. The reviews for factor $k_2$ are all about the quality and durability of the products, such as \textit{jammed shut} and \textit{solidly made}.
    \item By jointly analyzing the reviews across multiple factors, we find that, DGCLR is capable of modeling the interactions from multiple perspectives. In general, we character $k_1$, $k_2$, $k_3$, and $k_4$ as \textit{demand-supply match}, \textit{quality}, \textit{integrity} and \textit{cost performance}, respectively. This verifies our hypothesis that different uses' ratings on different items are driven by distinct latent factors.
\vspace{-0.1cm}
\end{itemize}
\vspace{-0.1cm}
\subsubsection{Interpretability of attention-based rating prediction.} 
Having interpreted the semantics of factor graphs, in this section, we provide explanations for the factor-level recommendation  based on our proposed AI module. Specifically, we randomly sample two user-item pairs with the highest attention weight $\alpha_k>0.5$ predicted by DGCLR from the test set of Office. Figure \ref{fig5} depicts the rating prediction process of the AI module, as well as the ground-truth rating scores and reviews. We can see that the AI module can identify the varied decisive factors $k_1$ and $k_4$ in the two interactions and make accurate (MSE<0.2) rating predictions $\hat{r}_{i,j}^{k}$
based on the decisive factor $k$. By comprehensively analyzing the two ground-truth reviews and the high-level concepts of factors $k_1$ and $k_4$ summarized in section 5.5.1, we discover, to our surprise, that they are semantically consistent, which demonstrates that DGCLR can capture the key factors that vary across user-item interactions.
\vspace{-0.2cm}
\section{Conclusion}
\vspace{-0.1cm}
This paper proposes a novel framework, DGCLR, which mainly focuses on exploring and disentangling the latent factors behind user-item interactions for better and more explainable review-based recommendation. Specifically, we design a disentangled graph learning module to factorize the user-item rating graph and  learn disentangled user/item representations. Then, an attention-based interaction module is created to model user-item interactions and make rating predictions adaptively from different latent factors. In addition, we introduce two factor-wise contrastive learning tasks to ensure that each factor representation is sufficiently discriminative and disentangled to reflect user/item and interaction properties. Experiments on benchmark datasets validate the effectiveness and interpretability of DGCLR.

\bibliographystyle{ACM-Reference-Format}
\bibliography{sample-base}


\begin{thebibliography}{52}


\ifx \showCODEN    \undefined \def \showCODEN     #1{\unskip}     \fi
\ifx \showDOI      \undefined \def \showDOI       #1{#1}\fi
\ifx \showISBNx    \undefined \def \showISBNx     #1{\unskip}     \fi
\ifx \showISBNxiii \undefined \def \showISBNxiii  #1{\unskip}     \fi
\ifx \showISSN     \undefined \def \showISSN      #1{\unskip}     \fi
\ifx \showLCCN     \undefined \def \showLCCN      #1{\unskip}     \fi
\ifx \shownote     \undefined \def \shownote      #1{#1}          \fi
\ifx \showarticletitle \undefined \def \showarticletitle #1{#1}   \fi
\ifx \showURL      \undefined \def \showURL       {\relax}        \fi
\providecommand\bibfield[2]{#2}
\providecommand\bibinfo[2]{#2}
\providecommand\natexlab[1]{#1}
\providecommand\showeprint[2][]{arXiv:#2}

\bibitem[\protect\citeauthoryear{Berg, Kipf, and Welling}{Berg
  et~al\mbox{.}}{2017}]%
        {berg2017graph}
\bibfield{author}{\bibinfo{person}{Rianne van~den Berg},
  \bibinfo{person}{Thomas~N Kipf}, {and} \bibinfo{person}{Max Welling}.}
  \bibinfo{year}{2017}\natexlab{}.
\newblock \showarticletitle{Graph convolutional matrix completion}.
\newblock \bibinfo{journal}{\emph{arXiv preprint arXiv:1706.02263}}
  (\bibinfo{year}{2017}).
\newblock


\bibitem[\protect\citeauthoryear{Blei, Ng, and Jordan}{Blei
  et~al\mbox{.}}{2003}]%
        {blei2003latent}
\bibfield{author}{\bibinfo{person}{David~M Blei}, \bibinfo{person}{Andrew~Y
  Ng}, {and} \bibinfo{person}{Michael~I Jordan}.}
  \bibinfo{year}{2003}\natexlab{}.
\newblock \showarticletitle{Latent dirichlet allocation}.
\newblock \bibinfo{journal}{\emph{JMLR}} \bibinfo{volume}{3},
  \bibinfo{number}{Jan} (\bibinfo{year}{2003}), \bibinfo{pages}{993--1022}.
\newblock


\bibitem[\protect\citeauthoryear{Catherine and Cohen}{Catherine and
  Cohen}{2017}]%
        {catherine2017transnets}
\bibfield{author}{\bibinfo{person}{Rose Catherine} {and}
  \bibinfo{person}{William Cohen}.} \bibinfo{year}{2017}\natexlab{}.
\newblock \showarticletitle{Transnets: Learning to transform for
  recommendation}. In \bibinfo{booktitle}{\emph{Proceedings of the eleventh ACM
  conference on recommender systems}}. \bibinfo{pages}{288--296}.
\newblock


\bibitem[\protect\citeauthoryear{Chen, Zhang, Liu, and Ma}{Chen
  et~al\mbox{.}}{2018}]%
        {chen2018neural}
\bibfield{author}{\bibinfo{person}{Chong Chen}, \bibinfo{person}{Min Zhang},
  \bibinfo{person}{Yiqun Liu}, {and} \bibinfo{person}{Shaoping Ma}.}
  \bibinfo{year}{2018}\natexlab{}.
\newblock \showarticletitle{Neural attentional rating regression with
  review-level explanations}. In \bibinfo{booktitle}{\emph{WWW}}.
  \bibinfo{pages}{1583--1592}.
\newblock


\bibitem[\protect\citeauthoryear{Chen, Chen, Wang, Xie, Wang, Xia, and
  Zhu}{Chen et~al\mbox{.}}{2021}]%
        {chen2021curriculum}
\bibfield{author}{\bibinfo{person}{Hong Chen}, \bibinfo{person}{Yudong Chen},
  \bibinfo{person}{Xin Wang}, \bibinfo{person}{Ruobing Xie},
  \bibinfo{person}{Rui Wang}, \bibinfo{person}{Feng Xia}, {and}
  \bibinfo{person}{Wenwu Zhu}.} \bibinfo{year}{2021}\natexlab{}.
\newblock \showarticletitle{Curriculum Disentangled Recommendation with Noisy
  Multi-feedback}.
\newblock \bibinfo{journal}{\emph{NeurIPS}}  \bibinfo{volume}{34}
  (\bibinfo{year}{2021}), \bibinfo{pages}{26924--26936}.
\newblock


\bibitem[\protect\citeauthoryear{Chen, Kornblith, Norouzi, and Hinton}{Chen
  et~al\mbox{.}}{2020}]%
        {chen2020simple}
\bibfield{author}{\bibinfo{person}{Ting Chen}, \bibinfo{person}{Simon
  Kornblith}, \bibinfo{person}{Mohammad Norouzi}, {and}
  \bibinfo{person}{Geoffrey Hinton}.} \bibinfo{year}{2020}\natexlab{}.
\newblock \showarticletitle{A simple framework for contrastive learning of
  visual representations}. In \bibinfo{booktitle}{\emph{International
  conference on machine learning}}. PMLR, \bibinfo{pages}{1597--1607}.
\newblock


\bibitem[\protect\citeauthoryear{Gao, Lin, Wang, Wang, Yang, He, and Chu}{Gao
  et~al\mbox{.}}{2020}]%
        {gao2020set}
\bibfield{author}{\bibinfo{person}{Jingyue Gao}, \bibinfo{person}{Yang Lin},
  \bibinfo{person}{Yasha Wang}, \bibinfo{person}{Xiting Wang},
  \bibinfo{person}{Zhao Yang}, \bibinfo{person}{Yuanduo He}, {and}
  \bibinfo{person}{Xu Chu}.} \bibinfo{year}{2020}\natexlab{}.
\newblock \showarticletitle{Set-sequence-graph: A multi-view approach towards
  exploiting reviews for recommendation}. In \bibinfo{booktitle}{\emph{CIKM}}.
  \bibinfo{pages}{395--404}.
\newblock


\bibitem[\protect\citeauthoryear{Glorot and Bengio}{Glorot and Bengio}{2010}]%
        {glorot2010understanding}
\bibfield{author}{\bibinfo{person}{Xavier Glorot} {and} \bibinfo{person}{Yoshua
  Bengio}.} \bibinfo{year}{2010}\natexlab{}.
\newblock \showarticletitle{Understanding the difficulty of training deep
  feedforward neural networks}. In \bibinfo{booktitle}{\emph{Proceedings of the
  thirteenth international conference on artificial intelligence and
  statistics}}. JMLR Workshop and Conference Proceedings,
  \bibinfo{pages}{249--256}.
\newblock


\bibitem[\protect\citeauthoryear{He and McAuley}{He and McAuley}{2016}]%
        {he2016ups}
\bibfield{author}{\bibinfo{person}{Ruining He} {and} \bibinfo{person}{Julian
  McAuley}.} \bibinfo{year}{2016}\natexlab{}.
\newblock \showarticletitle{Ups and downs: Modeling the visual evolution of
  fashion trends with one-class collaborative filtering}. In
  \bibinfo{booktitle}{\emph{WWW}}. \bibinfo{pages}{507--517}.
\newblock


\bibitem[\protect\citeauthoryear{He, Liao, Zhang, Nie, Hu, and Chua}{He
  et~al\mbox{.}}{2017}]%
        {he2017neural}
\bibfield{author}{\bibinfo{person}{Xiangnan He}, \bibinfo{person}{Lizi Liao},
  \bibinfo{person}{Hanwang Zhang}, \bibinfo{person}{Liqiang Nie},
  \bibinfo{person}{Xia Hu}, {and} \bibinfo{person}{Tat-Seng Chua}.}
  \bibinfo{year}{2017}\natexlab{}.
\newblock \showarticletitle{Neural collaborative filtering}. In
  \bibinfo{booktitle}{\emph{WWW}}. \bibinfo{pages}{173--182}.
\newblock


\bibitem[\protect\citeauthoryear{Higgins, Matthey, Pal, Burgess, Glorot,
  Botvinick, Mohamed, and Lerchner}{Higgins et~al\mbox{.}}{2016}]%
        {higgins2016beta}
\bibfield{author}{\bibinfo{person}{Irina Higgins}, \bibinfo{person}{Loic
  Matthey}, \bibinfo{person}{Arka Pal}, \bibinfo{person}{Christopher Burgess},
  \bibinfo{person}{Xavier Glorot}, \bibinfo{person}{Matthew Botvinick},
  \bibinfo{person}{Shakir Mohamed}, {and} \bibinfo{person}{Alexander
  Lerchner}.} \bibinfo{year}{2016}\natexlab{}.
\newblock \showarticletitle{beta-vae: Learning basic visual concepts with a
  constrained variational framework}.
\newblock  (\bibinfo{year}{2016}).
\newblock


\bibitem[\protect\citeauthoryear{Hyun, Park, Yang, Song, Lee, and Yu}{Hyun
  et~al\mbox{.}}{2018}]%
        {hyun2018review}
\bibfield{author}{\bibinfo{person}{Dongmin Hyun}, \bibinfo{person}{Chanyoung
  Park}, \bibinfo{person}{Min-Chul Yang}, \bibinfo{person}{Ilhyeon Song},
  \bibinfo{person}{Jung-Tae Lee}, {and} \bibinfo{person}{Hwanjo Yu}.}
  \bibinfo{year}{2018}\natexlab{}.
\newblock \showarticletitle{Review sentiment-guided scalable deep recommender
  system}. In \bibinfo{booktitle}{\emph{SIGIR}}. \bibinfo{pages}{965--968}.
\newblock


\bibitem[\protect\citeauthoryear{Kim}{Kim}{2014}]%
        {2014Convolutional}
\bibfield{author}{\bibinfo{person}{Y. Kim}.} \bibinfo{year}{2014}\natexlab{}.
\newblock \showarticletitle{Convolutional Neural Networks for Sentence
  Classification}. In \bibinfo{booktitle}{\emph{EMNLP}}.
  \bibinfo{pages}{1746--1751}.
\newblock


\bibitem[\protect\citeauthoryear{Kingma and Ba}{Kingma and Ba}{2014}]%
        {kingma2014adam}
\bibfield{author}{\bibinfo{person}{Diederik~P Kingma} {and}
  \bibinfo{person}{Jimmy Ba}.} \bibinfo{year}{2014}\natexlab{}.
\newblock \showarticletitle{Adam: A method for stochastic optimization}.
\newblock \bibinfo{journal}{\emph{arXiv preprint arXiv:1412.6980}}
  (\bibinfo{year}{2014}).
\newblock


\bibitem[\protect\citeauthoryear{Kipf and Welling}{Kipf and Welling}{2016}]%
        {kipf2016semi}
\bibfield{author}{\bibinfo{person}{Thomas~N Kipf} {and} \bibinfo{person}{Max
  Welling}.} \bibinfo{year}{2016}\natexlab{}.
\newblock \showarticletitle{Semi-supervised classification with graph
  convolutional networks}.
\newblock \bibinfo{journal}{\emph{arXiv preprint arXiv:1609.02907}}
  (\bibinfo{year}{2016}).
\newblock


\bibitem[\protect\citeauthoryear{Koren, Bell, and Volinsky}{Koren
  et~al\mbox{.}}{2009}]%
        {koren2009matrix}
\bibfield{author}{\bibinfo{person}{Yehuda Koren}, \bibinfo{person}{Robert
  Bell}, {and} \bibinfo{person}{Chris Volinsky}.}
  \bibinfo{year}{2009}\natexlab{}.
\newblock \showarticletitle{Matrix factorization techniques for recommender
  systems}.
\newblock \bibinfo{journal}{\emph{Computer}} \bibinfo{volume}{42},
  \bibinfo{number}{8} (\bibinfo{year}{2009}), \bibinfo{pages}{30--37}.
\newblock


\bibitem[\protect\citeauthoryear{Koren, Rendle, and Bell}{Koren
  et~al\mbox{.}}{2022}]%
        {koren2022advances}
\bibfield{author}{\bibinfo{person}{Yehuda Koren}, \bibinfo{person}{Steffen
  Rendle}, {and} \bibinfo{person}{Robert Bell}.}
  \bibinfo{year}{2022}\natexlab{}.
\newblock \showarticletitle{Advances in collaborative filtering}.
\newblock \bibinfo{journal}{\emph{Recommender systems handbook}}
  (\bibinfo{year}{2022}), \bibinfo{pages}{91--142}.
\newblock


\bibitem[\protect\citeauthoryear{Li, Cheng, Kshetramade, Houser, Chen, and
  Wang}{Li et~al\mbox{.}}{2021}]%
        {li2021recommend}
\bibfield{author}{\bibinfo{person}{Zeyu Li}, \bibinfo{person}{Wei Cheng},
  \bibinfo{person}{Reema Kshetramade}, \bibinfo{person}{John Houser},
  \bibinfo{person}{Haifeng Chen}, {and} \bibinfo{person}{Wei Wang}.}
  \bibinfo{year}{2021}\natexlab{}.
\newblock \showarticletitle{Recommend for a Reason: Unlocking the Power of
  Unsupervised Aspect-Sentiment Co-Extraction}.
\newblock \bibinfo{journal}{\emph{arXiv preprint arXiv:2109.03821}}
  (\bibinfo{year}{2021}).
\newblock


\bibitem[\protect\citeauthoryear{Liu, Li, Du, Chang, and Gao}{Liu
  et~al\mbox{.}}{2019}]%
        {liu2019daml}
\bibfield{author}{\bibinfo{person}{Donghua Liu}, \bibinfo{person}{Jing Li},
  \bibinfo{person}{Bo Du}, \bibinfo{person}{Jun Chang}, {and}
  \bibinfo{person}{Rong Gao}.} \bibinfo{year}{2019}\natexlab{}.
\newblock \showarticletitle{Daml: Dual attention mutual learning between
  ratings and reviews for item recommendation}. In
  \bibinfo{booktitle}{\emph{SIGKDD}}. \bibinfo{pages}{344--352}.
\newblock


\bibitem[\protect\citeauthoryear{Liu, Yang, Zhang, Miao, Nie, and Zhang}{Liu
  et~al\mbox{.}}{2021}]%
        {liu2021learning}
\bibfield{author}{\bibinfo{person}{Yong Liu}, \bibinfo{person}{Susen Yang},
  \bibinfo{person}{Yinan Zhang}, \bibinfo{person}{Chunyan Miao},
  \bibinfo{person}{Zaiqing Nie}, {and} \bibinfo{person}{Juyong Zhang}.}
  \bibinfo{year}{2021}\natexlab{}.
\newblock \showarticletitle{Learning hierarchical review graph representations
  for recommendation}.
\newblock \bibinfo{journal}{\emph{IEEE Transactions on Knowledge and Data
  Engineering}} (\bibinfo{year}{2021}).
\newblock


\bibitem[\protect\citeauthoryear{Ma, Cui, Kuang, Wang, and Zhu}{Ma
  et~al\mbox{.}}{2019a}]%
        {ma2019disentangled}
\bibfield{author}{\bibinfo{person}{Jianxin Ma}, \bibinfo{person}{Peng Cui},
  \bibinfo{person}{Kun Kuang}, \bibinfo{person}{Xin Wang}, {and}
  \bibinfo{person}{Wenwu Zhu}.} \bibinfo{year}{2019}\natexlab{a}.
\newblock \showarticletitle{Disentangled graph convolutional networks}. In
  \bibinfo{booktitle}{\emph{ICML}}. PMLR, \bibinfo{pages}{4212--4221}.
\newblock


\bibitem[\protect\citeauthoryear{Ma, Zhou, Cui, Yang, and Zhu}{Ma
  et~al\mbox{.}}{2019b}]%
        {ma2019learning}
\bibfield{author}{\bibinfo{person}{Jianxin Ma}, \bibinfo{person}{Chang Zhou},
  \bibinfo{person}{Peng Cui}, \bibinfo{person}{Hongxia Yang}, {and}
  \bibinfo{person}{Wenwu Zhu}.} \bibinfo{year}{2019}\natexlab{b}.
\newblock \showarticletitle{Learning disentangled representations for
  recommendation}.
\newblock \bibinfo{journal}{\emph{NeurIPS}}  \bibinfo{volume}{32}
  (\bibinfo{year}{2019}).
\newblock


\bibitem[\protect\citeauthoryear{Ma, Zhou, Yang, Cui, Wang, and Zhu}{Ma
  et~al\mbox{.}}{2020}]%
        {ma2020disentangled}
\bibfield{author}{\bibinfo{person}{Jianxin Ma}, \bibinfo{person}{Chang Zhou},
  \bibinfo{person}{Hongxia Yang}, \bibinfo{person}{Peng Cui},
  \bibinfo{person}{Xin Wang}, {and} \bibinfo{person}{Wenwu Zhu}.}
  \bibinfo{year}{2020}\natexlab{}.
\newblock \showarticletitle{Disentangled self-supervision in sequential
  recommenders}. In \bibinfo{booktitle}{\emph{SIGKDD}}.
  \bibinfo{pages}{483--491}.
\newblock


\bibitem[\protect\citeauthoryear{Mao, Lu, Zhang, and Zhang}{Mao
  et~al\mbox{.}}{2016}]%
        {mao2016multirelational}
\bibfield{author}{\bibinfo{person}{Mingsong Mao}, \bibinfo{person}{Jie Lu},
  \bibinfo{person}{Guangquan Zhang}, {and} \bibinfo{person}{Jinlong Zhang}.}
  \bibinfo{year}{2016}\natexlab{}.
\newblock \showarticletitle{Multirelational social recommendations via
  multigraph ranking}.
\newblock \bibinfo{journal}{\emph{IEEE transactions on cybernetics}}
  \bibinfo{volume}{47}, \bibinfo{number}{12} (\bibinfo{year}{2016}),
  \bibinfo{pages}{4049--4061}.
\newblock


\bibitem[\protect\citeauthoryear{McAuley and Leskovec}{McAuley and
  Leskovec}{2013}]%
        {mcauley2013hidden}
\bibfield{author}{\bibinfo{person}{Julian McAuley} {and} \bibinfo{person}{Jure
  Leskovec}.} \bibinfo{year}{2013}\natexlab{}.
\newblock \showarticletitle{Hidden factors and hidden topics: understanding
  rating dimensions with review text}. In \bibinfo{booktitle}{\emph{Proceedings
  of the 7th ACM conference on Recommender systems}}.
  \bibinfo{pages}{165--172}.
\newblock


\bibitem[\protect\citeauthoryear{Mikolov, Chen, Corrado, and Dean}{Mikolov
  et~al\mbox{.}}{2013}]%
        {mikolov2013efficient}
\bibfield{author}{\bibinfo{person}{Tomas Mikolov}, \bibinfo{person}{Kai Chen},
  \bibinfo{person}{Greg Corrado}, {and} \bibinfo{person}{Jeffrey Dean}.}
  \bibinfo{year}{2013}\natexlab{}.
\newblock \showarticletitle{Efficient estimation of word representations in
  vector space}.
\newblock \bibinfo{journal}{\emph{arXiv preprint arXiv:1301.3781}}
  (\bibinfo{year}{2013}).
\newblock


\bibitem[\protect\citeauthoryear{Mnih and Salakhutdinov}{Mnih and
  Salakhutdinov}{2007}]%
        {mnih2007probabilistic}
\bibfield{author}{\bibinfo{person}{Andriy Mnih} {and} \bibinfo{person}{Russ~R
  Salakhutdinov}.} \bibinfo{year}{2007}\natexlab{}.
\newblock \showarticletitle{Probabilistic matrix factorization}.
\newblock \bibinfo{journal}{\emph{NeurIPS}}  \bibinfo{volume}{20}
  (\bibinfo{year}{2007}).
\newblock


\bibitem[\protect\citeauthoryear{Oord, Li, and Vinyals}{Oord
  et~al\mbox{.}}{2018}]%
        {oord2018representation}
\bibfield{author}{\bibinfo{person}{Aaron van~den Oord}, \bibinfo{person}{Yazhe
  Li}, {and} \bibinfo{person}{Oriol Vinyals}.} \bibinfo{year}{2018}\natexlab{}.
\newblock \showarticletitle{Representation learning with contrastive predictive
  coding}.
\newblock \bibinfo{journal}{\emph{arXiv preprint arXiv:1807.03748}}
  (\bibinfo{year}{2018}).
\newblock


\bibitem[\protect\citeauthoryear{Pe{\~n}a, O'Reilly-Morgan, Tragos, Hurley,
  Duriakova, Smyth, and Lawlor}{Pe{\~n}a et~al\mbox{.}}{2020}]%
        {pena2020combining}
\bibfield{author}{\bibinfo{person}{Francisco~J Pe{\~n}a},
  \bibinfo{person}{Diarmuid O'Reilly-Morgan}, \bibinfo{person}{Elias~Z Tragos},
  \bibinfo{person}{Neil Hurley}, \bibinfo{person}{Erika Duriakova},
  \bibinfo{person}{Barry Smyth}, {and} \bibinfo{person}{Aonghus Lawlor}.}
  \bibinfo{year}{2020}\natexlab{}.
\newblock \showarticletitle{Combining rating and review data by initializing
  latent factor models with topic models for top-n recommendation}. In
  \bibinfo{booktitle}{\emph{Fourteenth ACM conference on recommender systems}}.
  \bibinfo{pages}{438--443}.
\newblock


\bibitem[\protect\citeauthoryear{Rendle, Freudenthaler, Gantner, and
  Schmidt-Thieme}{Rendle et~al\mbox{.}}{2012}]%
        {rendle2012bpr}
\bibfield{author}{\bibinfo{person}{Steffen Rendle}, \bibinfo{person}{Christoph
  Freudenthaler}, \bibinfo{person}{Zeno Gantner}, {and} \bibinfo{person}{Lars
  Schmidt-Thieme}.} \bibinfo{year}{2012}\natexlab{}.
\newblock \showarticletitle{BPR: Bayesian personalized ranking from implicit
  feedback}.
\newblock \bibinfo{journal}{\emph{arXiv preprint arXiv:1205.2618}}
  (\bibinfo{year}{2012}).
\newblock


\bibitem[\protect\citeauthoryear{Shi, Han, Song, Wang, Wang, Du, and
  Philip}{Shi et~al\mbox{.}}{2019}]%
        {shi2019deep}
\bibfield{author}{\bibinfo{person}{Chuan Shi}, \bibinfo{person}{Xiaotian Han},
  \bibinfo{person}{Li Song}, \bibinfo{person}{Xiao Wang},
  \bibinfo{person}{Senzhang Wang}, \bibinfo{person}{Junping Du}, {and}
  \bibinfo{person}{S~Yu Philip}.} \bibinfo{year}{2019}\natexlab{}.
\newblock \showarticletitle{Deep collaborative filtering with multi-aspect
  information in heterogeneous networks}.
\newblock \bibinfo{journal}{\emph{IEEE transactions on knowledge and data
  engineering}} \bibinfo{volume}{33}, \bibinfo{number}{4}
  (\bibinfo{year}{2019}), \bibinfo{pages}{1413--1425}.
\newblock


\bibitem[\protect\citeauthoryear{Shuai, Zhang, Wu, Sun, Hong, Wang, and
  Li}{Shuai et~al\mbox{.}}{2022}]%
        {shuai2022review}
\bibfield{author}{\bibinfo{person}{Jie Shuai}, \bibinfo{person}{Kun Zhang},
  \bibinfo{person}{Le Wu}, \bibinfo{person}{Peijie Sun},
  \bibinfo{person}{Richang Hong}, \bibinfo{person}{Meng Wang}, {and}
  \bibinfo{person}{Yong Li}.} \bibinfo{year}{2022}\natexlab{}.
\newblock \showarticletitle{A Review-aware Graph Contrastive Learning Framework
  for Recommendation}.
\newblock \bibinfo{journal}{\emph{arXiv preprint arXiv:2204.12063}}
  (\bibinfo{year}{2022}).
\newblock


\bibitem[\protect\citeauthoryear{Su, Cao, Liu, and Ou}{Su
  et~al\mbox{.}}{2021}]%
        {su2021whitening}
\bibfield{author}{\bibinfo{person}{Jianlin Su}, \bibinfo{person}{Jiarun Cao},
  \bibinfo{person}{Weijie Liu}, {and} \bibinfo{person}{Yangyiwen Ou}.}
  \bibinfo{year}{2021}\natexlab{}.
\newblock \showarticletitle{Whitening sentence representations for better
  semantics and faster retrieval}.
\newblock \bibinfo{journal}{\emph{arXiv preprint arXiv:2103.15316}}
  (\bibinfo{year}{2021}).
\newblock


\bibitem[\protect\citeauthoryear{Sun, Wu, Zhang, Fu, Hong, and Wang}{Sun
  et~al\mbox{.}}{2020}]%
        {sun2020dual}
\bibfield{author}{\bibinfo{person}{Peijie Sun}, \bibinfo{person}{Le Wu},
  \bibinfo{person}{Kun Zhang}, \bibinfo{person}{Yanjie Fu},
  \bibinfo{person}{Richang Hong}, {and} \bibinfo{person}{Meng Wang}.}
  \bibinfo{year}{2020}\natexlab{}.
\newblock \showarticletitle{Dual learning for explainable recommendation:
  Towards unifying user preference prediction and review generation}. In
  \bibinfo{booktitle}{\emph{WWW}}. \bibinfo{pages}{837--847}.
\newblock


\bibitem[\protect\citeauthoryear{Tay, Luu, and Hui}{Tay et~al\mbox{.}}{2018}]%
        {tay2018multi}
\bibfield{author}{\bibinfo{person}{Yi Tay}, \bibinfo{person}{Anh~Tuan Luu},
  {and} \bibinfo{person}{Siu~Cheung Hui}.} \bibinfo{year}{2018}\natexlab{}.
\newblock \showarticletitle{Multi-pointer co-attention networks for
  recommendation}. In \bibinfo{booktitle}{\emph{SIGKDD}}.
  \bibinfo{pages}{2309--2318}.
\newblock


\bibitem[\protect\citeauthoryear{Vaswani, Shazeer, Parmar, Uszkoreit, Jones,
  Gomez, Kaiser, and Polosukhin}{Vaswani et~al\mbox{.}}{2017}]%
        {vaswani2017attention}
\bibfield{author}{\bibinfo{person}{Ashish Vaswani}, \bibinfo{person}{Noam
  Shazeer}, \bibinfo{person}{Niki Parmar}, \bibinfo{person}{Jakob Uszkoreit},
  \bibinfo{person}{Llion Jones}, \bibinfo{person}{Aidan~N Gomez},
  \bibinfo{person}{{\L}ukasz Kaiser}, {and} \bibinfo{person}{Illia
  Polosukhin}.} \bibinfo{year}{2017}\natexlab{}.
\newblock \showarticletitle{Attention is all you need}.
\newblock \bibinfo{journal}{\emph{NeurIPS}}  \bibinfo{volume}{30}
  (\bibinfo{year}{2017}).
\newblock


\bibitem[\protect\citeauthoryear{Veli{\v{c}}kovi{\'c}, Cucurull, Casanova,
  Romero, Lio, and Bengio}{Veli{\v{c}}kovi{\'c} et~al\mbox{.}}{2017}]%
        {velivckovic2017graph}
\bibfield{author}{\bibinfo{person}{Petar Veli{\v{c}}kovi{\'c}},
  \bibinfo{person}{Guillem Cucurull}, \bibinfo{person}{Arantxa Casanova},
  \bibinfo{person}{Adriana Romero}, \bibinfo{person}{Pietro Lio}, {and}
  \bibinfo{person}{Yoshua Bengio}.} \bibinfo{year}{2017}\natexlab{}.
\newblock \showarticletitle{Graph attention networks}.
\newblock \bibinfo{journal}{\emph{arXiv preprint arXiv:1710.10903}}
  (\bibinfo{year}{2017}).
\newblock


\bibitem[\protect\citeauthoryear{Wang and Blei}{Wang and Blei}{2011}]%
        {wang2011collaborative}
\bibfield{author}{\bibinfo{person}{Chong Wang} {and} \bibinfo{person}{David~M
  Blei}.} \bibinfo{year}{2011}\natexlab{}.
\newblock \showarticletitle{Collaborative topic modeling for recommending
  scientific articles}. In \bibinfo{booktitle}{\emph{SIGKDD}}.
  \bibinfo{pages}{448--456}.
\newblock


\bibitem[\protect\citeauthoryear{Wang, Wang, and Yeung}{Wang
  et~al\mbox{.}}{2015}]%
        {wang2015collaborative}
\bibfield{author}{\bibinfo{person}{Hao Wang}, \bibinfo{person}{Naiyan Wang},
  {and} \bibinfo{person}{Dit-Yan Yeung}.} \bibinfo{year}{2015}\natexlab{}.
\newblock \showarticletitle{Collaborative deep learning for recommender
  systems}. In \bibinfo{booktitle}{\emph{SIGKDD}}. \bibinfo{pages}{1235--1244}.
\newblock


\bibitem[\protect\citeauthoryear{Wang, Jin, Zhang, He, Xu, and Chua}{Wang
  et~al\mbox{.}}{2020}]%
        {wang2020disentangled}
\bibfield{author}{\bibinfo{person}{Xiang Wang}, \bibinfo{person}{Hongye Jin},
  \bibinfo{person}{An Zhang}, \bibinfo{person}{Xiangnan He},
  \bibinfo{person}{Tong Xu}, {and} \bibinfo{person}{Tat-Seng Chua}.}
  \bibinfo{year}{2020}\natexlab{}.
\newblock \showarticletitle{Disentangled graph collaborative filtering}. In
  \bibinfo{booktitle}{\emph{SIGIR}}. \bibinfo{pages}{1001--1010}.
\newblock


\bibitem[\protect\citeauthoryear{Wu, Wu, Liu, and Huang}{Wu
  et~al\mbox{.}}{2019b}]%
        {wu2019hierarchical}
\bibfield{author}{\bibinfo{person}{Chuhan Wu}, \bibinfo{person}{Fangzhao Wu},
  \bibinfo{person}{Junxin Liu}, {and} \bibinfo{person}{Yongfeng Huang}.}
  \bibinfo{year}{2019}\natexlab{b}.
\newblock \showarticletitle{Hierarchical user and item representation with
  three-tier attention for recommendation}. In
  \bibinfo{booktitle}{\emph{Proceedings of the 2019 Conference of the North
  American Chapter of the Association for Computational Linguistics: Human
  Language Technologies, Volume 1 (Long and Short Papers)}}.
  \bibinfo{pages}{1818--1826}.
\newblock


\bibitem[\protect\citeauthoryear{Wu, Wu, Qi, Ge, Huang, and Xie}{Wu
  et~al\mbox{.}}{2019c}]%
        {wu2019reviews}
\bibfield{author}{\bibinfo{person}{Chuhan Wu}, \bibinfo{person}{Fangzhao Wu},
  \bibinfo{person}{Tao Qi}, \bibinfo{person}{Suyu Ge},
  \bibinfo{person}{Yongfeng Huang}, {and} \bibinfo{person}{Xing Xie}.}
  \bibinfo{year}{2019}\natexlab{c}.
\newblock \showarticletitle{Reviews meet graphs: enhancing user and item
  representations for recommendation with hierarchical attentive graph neural
  network}. In \bibinfo{booktitle}{\emph{EMNLP}}. \bibinfo{pages}{4884--4893}.
\newblock


\bibitem[\protect\citeauthoryear{Wu, Wang, Feng, He, Chen, Lian, and Xie}{Wu
  et~al\mbox{.}}{2021}]%
        {wu2021self}
\bibfield{author}{\bibinfo{person}{Jiancan Wu}, \bibinfo{person}{Xiang Wang},
  \bibinfo{person}{Fuli Feng}, \bibinfo{person}{Xiangnan He},
  \bibinfo{person}{Liang Chen}, \bibinfo{person}{Jianxun Lian}, {and}
  \bibinfo{person}{Xing Xie}.} \bibinfo{year}{2021}\natexlab{}.
\newblock \showarticletitle{Self-supervised graph learning for recommendation}.
  In \bibinfo{booktitle}{\emph{SIGIR}}. \bibinfo{pages}{726--735}.
\newblock


\bibitem[\protect\citeauthoryear{Wu, Quan, Li, Wang, Zheng, and Luo}{Wu
  et~al\mbox{.}}{2019a}]%
        {wu2019context}
\bibfield{author}{\bibinfo{person}{Libing Wu}, \bibinfo{person}{Cong Quan},
  \bibinfo{person}{Chenliang Li}, \bibinfo{person}{Qian Wang},
  \bibinfo{person}{Bolong Zheng}, {and} \bibinfo{person}{Xiangyang Luo}.}
  \bibinfo{year}{2019}\natexlab{a}.
\newblock \showarticletitle{A context-aware user-item representation learning
  for item recommendation}.
\newblock \bibinfo{journal}{\emph{TOIS}} \bibinfo{volume}{37},
  \bibinfo{number}{2} (\bibinfo{year}{2019}), \bibinfo{pages}{1--29}.
\newblock


\bibitem[\protect\citeauthoryear{Xi, Huang, Wang, Zheng, and Lai}{Xi
  et~al\mbox{.}}{2021}]%
        {xi2021deep}
\bibfield{author}{\bibinfo{person}{Wu-Dong Xi}, \bibinfo{person}{Ling Huang},
  \bibinfo{person}{Chang-Dong Wang}, \bibinfo{person}{Yin-Yu Zheng}, {and}
  \bibinfo{person}{Jian-Huang Lai}.} \bibinfo{year}{2021}\natexlab{}.
\newblock \showarticletitle{Deep rating and review neural network for item
  recommendation}.
\newblock \bibinfo{journal}{\emph{TNNLS}} (\bibinfo{year}{2021}).
\newblock


\bibitem[\protect\citeauthoryear{Xia, Huang, Xu, Zhao, Yin, and Huang}{Xia
  et~al\mbox{.}}{2022}]%
        {xia2022hypergraph}
\bibfield{author}{\bibinfo{person}{Lianghao Xia}, \bibinfo{person}{Chao Huang},
  \bibinfo{person}{Yong Xu}, \bibinfo{person}{Jiashu Zhao},
  \bibinfo{person}{Dawei Yin}, {and} \bibinfo{person}{Jimmy Huang}.}
  \bibinfo{year}{2022}\natexlab{}.
\newblock \showarticletitle{Hypergraph contrastive collaborative filtering}. In
  \bibinfo{booktitle}{\emph{Proceedings of the 45th International ACM SIGIR
  Conference on Research and Development in Information Retrieval}}.
  \bibinfo{pages}{70--79}.
\newblock


\bibitem[\protect\citeauthoryear{Yang, Feng, Song, and Wang}{Yang
  et~al\mbox{.}}{2020}]%
        {yang2020factorizable}
\bibfield{author}{\bibinfo{person}{Yiding Yang}, \bibinfo{person}{Zunlei Feng},
  \bibinfo{person}{Mingli Song}, {and} \bibinfo{person}{Xinchao Wang}.}
  \bibinfo{year}{2020}\natexlab{}.
\newblock \showarticletitle{Factorizable graph convolutional networks}.
\newblock \bibinfo{journal}{\emph{NeurIPS}}  \bibinfo{volume}{33}
  (\bibinfo{year}{2020}), \bibinfo{pages}{20286--20296}.
\newblock


\bibitem[\protect\citeauthoryear{You, Chen, Sui, Chen, Wang, and Shen}{You
  et~al\mbox{.}}{2020}]%
        {you2020graph}
\bibfield{author}{\bibinfo{person}{Yuning You}, \bibinfo{person}{Tianlong
  Chen}, \bibinfo{person}{Yongduo Sui}, \bibinfo{person}{Ting Chen},
  \bibinfo{person}{Zhangyang Wang}, {and} \bibinfo{person}{Yang Shen}.}
  \bibinfo{year}{2020}\natexlab{}.
\newblock \showarticletitle{Graph contrastive learning with augmentations}.
\newblock \bibinfo{journal}{\emph{NeurIPS}}  \bibinfo{volume}{33}
  (\bibinfo{year}{2020}), \bibinfo{pages}{5812--5823}.
\newblock


\bibitem[\protect\citeauthoryear{Yu, Gao, Li, Yin, and Liu}{Yu
  et~al\mbox{.}}{2018}]%
        {yu2018adaptive}
\bibfield{author}{\bibinfo{person}{Junliang Yu}, \bibinfo{person}{Min Gao},
  \bibinfo{person}{Jundong Li}, \bibinfo{person}{Hongzhi Yin}, {and}
  \bibinfo{person}{Huan Liu}.} \bibinfo{year}{2018}\natexlab{}.
\newblock \showarticletitle{Adaptive implicit friends identification over
  heterogeneous network for social recommendation}. In
  \bibinfo{booktitle}{\emph{CIKM}}. \bibinfo{pages}{357--366}.
\newblock


\bibitem[\protect\citeauthoryear{Yu, Yin, Gao, Xia, Zhang, and Viet~Hung}{Yu
  et~al\mbox{.}}{2021}]%
        {yu2021socially}
\bibfield{author}{\bibinfo{person}{Junliang Yu}, \bibinfo{person}{Hongzhi Yin},
  \bibinfo{person}{Min Gao}, \bibinfo{person}{Xin Xia},
  \bibinfo{person}{Xiangliang Zhang}, {and} \bibinfo{person}{Nguyen~Quoc
  Viet~Hung}.} \bibinfo{year}{2021}\natexlab{}.
\newblock \showarticletitle{Socially-aware self-supervised tri-training for
  recommendation}. In \bibinfo{booktitle}{\emph{SIGKDD}}.
  \bibinfo{pages}{2084--2092}.
\newblock


\bibitem[\protect\citeauthoryear{Zhang, Gao, Yu, Guo, Li, and Yin}{Zhang
  et~al\mbox{.}}{2021}]%
        {zhang2021double}
\bibfield{author}{\bibinfo{person}{Junwei Zhang}, \bibinfo{person}{Min Gao},
  \bibinfo{person}{Junliang Yu}, \bibinfo{person}{Lei Guo},
  \bibinfo{person}{Jundong Li}, {and} \bibinfo{person}{Hongzhi Yin}.}
  \bibinfo{year}{2021}\natexlab{}.
\newblock \showarticletitle{Double-scale self-supervised hypergraph learning
  for group recommendation}. In \bibinfo{booktitle}{\emph{CIKM}}.
  \bibinfo{pages}{2557--2567}.
\newblock


\bibitem[\protect\citeauthoryear{Zheng, Noroozi, and Yu}{Zheng
  et~al\mbox{.}}{2017}]%
        {zheng2017joint}
\bibfield{author}{\bibinfo{person}{Lei Zheng}, \bibinfo{person}{Vahid Noroozi},
  {and} \bibinfo{person}{Philip~S Yu}.} \bibinfo{year}{2017}\natexlab{}.
\newblock \showarticletitle{Joint deep modeling of users and items using
  reviews for recommendation}. In \bibinfo{booktitle}{\emph{WSDM}}.
  \bibinfo{pages}{425--434}.
\newblock


\end{thebibliography}
\end{document}